\documentclass[11pt]{article}
\usepackage{geometry}                % See geometry.pdf to learn the layout options. There are lots.
\usepackage{graphicx}
\usepackage{epsfig}
\usepackage{color}
\usepackage{subfigure}
\usepackage{graphics}

\usepackage{amssymb}
\usepackage{amsthm}
\usepackage{amsmath}
\usepackage{amscd}
\usepackage{epstopdf}

\usepackage[ruled,longend]{algorithm2e}

\usepackage{fullpage}

\usepackage[ruled]{algorithm2e}

\newtheorem{theorem}{Theorem}[section]
\newtheorem{lemma}[theorem]{Lemma}
\newtheorem{proposition}[theorem]{Proposition}
\newtheorem{corollary}[theorem]{Corollary}

\newtheorem{definition}[theorem]{Definition}

\newtheorem{remark}[theorem]{Remark}
\newtheorem{observation}[theorem]{Observation}

\def\V{{\mathcal V}}   %alphabet of endpoints
\def\G{{Gr}} %accessibility graph function

\title{Practical and Efficient Circle Graph Recognition}
\author{
 
Emeric Gioan\footnotemark[1], Christophe Paul\footnotemark[1], Marc Tedder\footnotemark[2],  Derek Corneil\footnotemark[2]
}

\date{}                                           % Activate to display a given date or no date

\begin{document}
\maketitle

\footnotetext[1]{CNRS - LIRMM, Univ. Montpellier II France; \{emeric.gioan,christophe.paul\}@lirmm.fr; financial support was received from the French ANR project ANR-O6-BLAN-0148-01: \emph{Graph Decomposition and Algorithms} (GRAAL).}
\footnotetext[2]{Department of Computer Science, University of Toronto; \{mtedder,dgc\}@cs.toronto.edu; financial support was received from Canada's Natural Sciences and Engineering Research Council (NSERC).}
\addtocounter{footnote}{2}

\begin{abstract}
Circle graphs are the intersection graphs of chords in a circle.  This paper presents the first sub-quadratic recognition algorithm for the class of circle graphs.  Our algorithm is $O(n+m)$ times the inverse Ackermann function, $\alpha(n+m)$, whose value is smaller than 4 for any practical graph. The algorithm is based on a new incremental Lexicographic Breadth-First Search characterization of circle graphs, and a new efficient data-structure for circle graphs, both developed in the paper.  The algorithm is an extension of a Split Decomposition algorithm with the same running time developed by the authors in a companion paper.
% MT 08/28/12 - Should we put a reference here as we do in the split paper?  I don't like references in Abstracts, but we should at least be consistent.
% XTOF 09/2012 : I don't mind to not put a ref in the abstract. The companion paper is referred at other places.

% It uses split decomposition to verify that the conditions of the characterization are met, and more specifically, the recent reformulation of split decomposition in terms of graph-labelled trees.  
%A new data-structure for circle graphs developed here facilitates the algorithm's efficient implementation.  
\end{abstract}

%----------------------------------------------------------------------------------------------------------------------
%----------------------------------------------------------------------------------------------------------------------
\section{Introduction}

%A \emph{chord diagram} is a circle inscribed by a set of chords.  
A \emph{chord diagram} can be defined as a circle inscribed by a set of chords.  
%Eme-v7 I add "can be defines as" so that this def does not interefer with the later one, cf Marc's comments in Section 2.4
A graph is a \emph{circle graph} if it is the intersection graph 
of a chord diagram: the vertices correspond to the chords, and two vertices are adjacent if and only if their chords intersect.  
%EME
Combinatorially, chord diagrams are 
%represented 
defined
%Eme-v7 continuing remark above. Note that the topological definition is not correct to say that a prime graph has a unique chord diagram unless we define also a toplogical equivalence... that is why we don't care about the topological def and just consider the combinatorial one
by double occurence circular words.
Circle graphs were first introduced in the early 1970s, under the name \emph{alternance graphs}, as a means of sorting permutations using stacks~\cite{ET71}.  The polynomial time recognition of circle graphs was posed as an open problem by Golumbic in the first edition of his book~\cite{Gol04}.  The question received considerable attention afterwards and was eventually settled independently by Naji~\cite{Naj85}, Bouchet~\cite{Bou87}, and Gabor et al.~\cite{GHS89}. 

Bouchet's $O(n^5)$ algorithm is based on a characterization of circle graphs in terms of local complementation, a concept originated in his work on isotropic systems~\cite{Bou88}, of which the recently introduced rank-width and vertex-minor theories are extensions~\cite{Oum05,GO09}.  It is conjectured that circle graphs are related to rank-width and vertex-minors as planar graphs are related to tree-width and graph-minors: just as large tree-width implies the existence of a large grid as a graph-minor, it is conjectured that large rank-width implies the existence of a large circle graph vertex-minor~\cite{Oum09}.  The conjecture has already been verified for line-graphs~\cite{Oum09}.

Both Naji's $O(n^7)$ algorithm and Gabor et al.'s $O(n^3)$ algorithm are based on split decomposition, introduced by Cunningham~\cite{Cun82}.  A \emph{split} is a bipartition $(A,B)$ (with $|A|,|B|>1$) of a graph's vertices, where there are subsets (called the \emph{frontiers}) $A' \subseteq A$ and $B'\subseteq B$ such that no edges exist between $A$ and $B$ other than those between $A'$ and $B'$, and every possible edge exists between $A'$ and $B'$.  
%DGC:  added "called the frontiers" and removed the later incorrect definition of SD.
%Marc-v6: I don't think this is needed here
%The sets $A'$ and $B'$ are called the \emph{frontiers} of the split.  
%
Intuitively, split decomposition finds a split and recursively decomposes its parts.  A graph is called \emph{prime} if it does not contain a split.  It is known that a graph is a circle graph if and only if its prime split decomposition components are circle graphs~\cite{GHS89}. This property is used by 
%Naji as well as Gabor et al.
%Marc-v6: I think Bouchet did as well
Bouchet, Naji, and Gabor et al. to reduce the recognition of circle graphs to the recognition of prime circle graphs.  
The latter problem is made somewhat easier by the fact that prime circle graphs have unique chord diagrams (up to reflection)~\cite{Bou87} (see also \cite{Cou08}).    

The algorithm of Gabor et al. was improved by Spinrad in 1994 to run in time $O(n^2)$~\cite{Spi94}.   A key component is an $O(n^2)$ prime testing procedure he developed with Ma~\cite{MS94}.  A linear time prime testing procedure now exists in the form of Dahlhaus' %complicated 
split decomposition algorithm~\cite{Dah00}; however, a faster circle graph recognition algorithm has not followed.  
%The complexity bottleneck in Spinrad's algorithm is actually his procedure to construct the unique realizer for prime circle graphs.  The complexity of circle graph recognition had remained $O(n^2)$ until now.
%Marc-v6: for clarity
In fact, the complexity bottleneck in Spinrad's algorithm is not computing the split decomposition, but rather his procedure to construct the unique 
%realizer 
chord diagram
for prime circle graphs.  
%Eme-v7 I changed "realizer" into "chord diagram" above
%
%Marc-v6: why the skip?
%Eme-v7: it was just to separate old stuff from our new stuff
%\medskip

This paper presents the first sub-quadratic circle graph recognition algorithm.  Our algorithm runs in time $O(n+m) \alpha(n+m)$, where $\alpha$ is the 
inverse Ackermann function \cite{CLR01,Tar75}.
%functional inverse of Ackermann's function $k\mapsto A(k,k)$.
%EME
%We point out that this function is so slowly growing that it is bounded by $4$ for all practical purpose, precisely for all 
%Marc-v6
We point out that this function is so slowly growing that it is bounded by $4$ for all practical purposes%
\footnote{
\font\sixrm=cmr10 scaled 600
% MT 08/31/12 - Reviewer noted a typo or latex error but I could not find one.  Instead, I cleaned up the English.
 Let us  mention that several definitions exist for this function, either with two variables, including some variants, or with one variable. For simplicity, we choose to use the version with one variable. This makes no practical difference since all of them could be used in our complexity bound, and they are all essentially constant. As an example, the two variable function considered in \cite{CLR01} satisifies $\alpha(k,n)\leq 4$ for all integer $k$ and for all $n\leq \underbrace{2^{.^{.^{.^{2}}}}}_{17 \hbox{\sixrm { times}}}$.
}%
.

We overcome Spinrad's bottleneck in two ways: 
we use the recent reformulation of split decomposition in terms of graph-labelled trees (GLTs)~\cite{GP07,GP08}, and we derive a new characterization of circle graphs in terms of Lexicographic Breadth-First Search (LBFS)~\cite{RTL76}. 
%Marc-v6
The key technical concept we deal with is that of \emph{consecutiveness} in a chord diagram (Section \ref{sec:consecutivity}), a property that can be efficiently preserved under  a certain GLT transformation (Section \ref{sec:circle-join-preserving}).
%DGC "a" certain transformation.
%
%, a property that can be efficiently preserved under some conditions (Section \ref{sec:circle-join-preserving}).
%, a property that can be preserved while making our main elementary operation (Section \ref{sec:circle-join-preserving}).
On one hand, this concept provides a new property for 
%Marc-v6: consistency with above terminology
chord diagrams of the components
 in the split decomposition of a circle graph (Section \ref{sec:circle+split}).   
%Marc-v6
%On the other hand, it allows us to characterize how a prime circle graph can be built incrementally, according to an LBFS ordering (Section \ref{sec:good}).
%
%Eme-v7: modif to sum up each subsection better
%DGC  below and elsewhere "w.r.t." written out.
On the other hand, it provides a new property for prime circle graphs with respect to an LBFS ordering (Section \ref{sec:good}). Finally, these results allow us to characterize how a prime circle graph can be built incrementally, according to an LBFS ordering (Section \ref{sec:charact}).

%This treatment of prime circle graphs can be integrated in the incremental split decomposition algorithm from the companion paper~\cite{GPTC}, whose cost is $O(n+m) \alpha(n+m)$.  That algorithm operates in the GLT setting, computing the split decomposition incrementally, only it adds vertices according to an LBFS ordering. 
%Marc-v6
This treatment of prime circle graphs can be integrated with the incremental split decomposition algorithm from the companion paper~\cite{GPTC11a}, whose running time is $O(n+m) \alpha(n+m)$.  That algorithm operates in the GLT setting, computing the split decomposition incrementally, only it adds vertices according to an LBFS ordering %
%Eme-v7 added \ref to complete the summary
(Section \ref{sec:algorithm})%
.  
%
%So we essentially need to refine this algorithm by maintaining chord diagram layouts discussed above at the same time prime graphs are maintained (Section \ref{sec:algorithm}).  To this aim, a new data-structure we develop for chord diagrams permits to perform efficiently the required operations (Section \ref{sec:implementation+DS}).  We point out that our algorithm also gives the (unique) chord diagrams for all prime labels in the split decomposition of a circle graph.
%Marc-v6: clarify to help the reader
Throughout that process, our proposed circle graph recognition algorithm maintains chord diagrams for all prime components in the split decomposition so long as possible.  We do so by applying the new results mentioned above for prime circle graphs in an incremental LBFS setting.  
A new data-structure for chord diagrams is developed in the paper so that these results can be efficiently implemented %
%
%Eme-v7
(Section \ref{sec:implementation+DS})%
.  
In particular, our new data-structure is what enables the efficiency of the GLT transformations that preserve consecutiveness.  Our results represent substantial progress on a long-standing open problem.  
%

%----------------------------------------------------------------------------------------------------------------------
%----------------------------------------------------------------------------------------------------------------------

%Marc-v6: I have not checked that we use all of these.  I suspect that we don't given all the various drafts this has undergone.  Please verify.
%Eme-v7: it has been Christophe's job to verify this... hope he did it well !

\section{Preliminaries}

%----------------------------------------------------------------------------------------------------------------------
%Marc-v6: Word choice wrong given English usage.
%\subsection{Generalities - Classical notations}
\subsection{Basic Definitions and Terminology}
\label{sub:prel1}

All graphs in this document are simple, undirected, and connected.  The set of vertices in the graph $G$ is denoted $V(G)$ (or $V$ when the context is clear).  The subgraph  of $G$ \emph{induced} on the set of vertices $S$ is signified by $G[S]$. We let $N_G(x)$, or simply $N(x)$, denote the set of neighbours of $x$, and 
% Marc-v6 
if $S$ is a set of vertices, then $N(S)=(\cup_{x\in S} N(x))\setminus S$.  
A vertex is \emph{universal} to a set of vertices $S$ if it is adjacent to every vertex in $S$.
%DGC removed - it is \emph{isolated} from $S$ if it is adjacent to no vertex in $S$.  
A vertex is \emph{universal} in a graph if it is adjacent to every other vertex in the graph. A \emph{clique} is a graph in which every pair of vertices is adjacent.  We require in this paper that cliques have at least three vertices.  A \emph{star} is a graph with at least three vertices in which one vertex, called its \emph{centre}, is universal, and no other edges exist. 
%DGC  removed - The clique on $n$ vertices is denoted $K_n$; the star on $n$ vertices is denoted $S_n$.  
%The following text is moved up and the incorrect definition of SD is removed.
Cliques and stars are called \emph{degenerate} with respect to split decomposition as every non-trivial bipartition of their vertices forms a split.  Given two connected graphs $G$ and $G'$, each having at least two vertices, and given two vertices $q\in V(G)$ and $q'\in V(G')$, the \emph{join} between $G$ and $G'$ with respect to $q$ and $q'$, denoted by  $(G,q)\otimes (G',q')$, is the graph formed from $G$ and $G'$ as follows: all possible edges are added between $N_{G}(q)$ and 
%Marc-v6
$N_{G'}(q')$,
and then $q$ and $q'$ are deleted. In this case, observe that $(V(G)\setminus\{q\}, V(G')\setminus\{q'\})$ is a split of the graph  $(G,q)\otimes (G',q')$.

The graph $G+(x,N(x))$ is formed by adding the vertex $x$ to the graph $G$ adjacent to the subset $N(x)$ of vertices, its neighbourhood; when $N(x)$ is clear from the context we simply write $G+x$.  The graph $G-x$ is formed from $G$ by removing $x$ and all its incident edges.

To avoid confusion with graphs, the edges of a tree are called \emph{tree-edges}. If $T$ is a tree, then $|T|$ represents 
%Marc-v6
the number of its vertices.  
The non-leaf vertices of a tree are called its \emph{nodes}.  The tree-edges not incident to leaves are \emph{internal tree-edges}.  

\subsection{The Split-Tree of a Graph} \label{sec:GLT}
\label{sub:prel2}

The split decomposition and the 
%Marc-v6
related
split-tree play a central role in the circle graph recognition problem.
This subsection essentially recalls definitions from \cite{GP07, GP08} and from \cite{GPTC11a}. Here, we will give only the material required in the present paper. 
%More involved definitions and details are given in \cite{GPTC11a}. 
More involved definitions and details are given in \cite{GPTC11a}.
Let us mention that the graph-labelled tree sructure defined below can be easily related to other representations used for the split decomposition, e.g. \cite{Cou06,Cun82}.

\begin{figure}[h]    %[htbh]
\begin{center}
\includegraphics[scale=0.75]{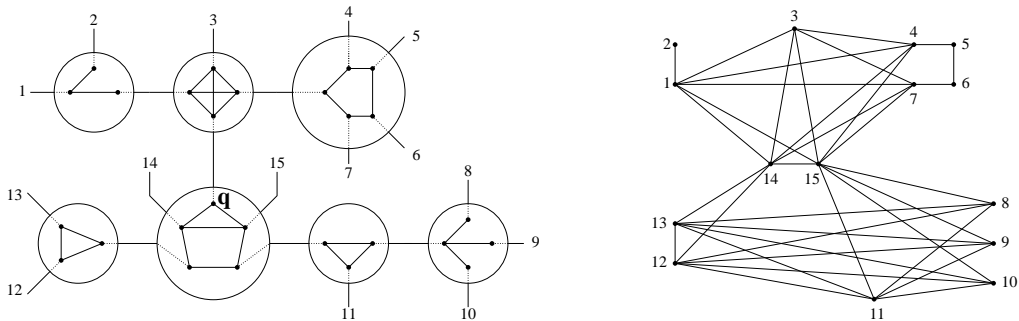}
\end{center}
\vspace{-0.5cm}
%Marc-v6: Fix to be clearer.
\caption{On the left: a graph-labelled tree $(T,\mathcal{F})$; on the right: its accessibility graph $\G(T,\mathcal{F})$. For the pictured marker vertex $q$, we have $L(q)=\{1,2,3,4,5,6,7\}$. The leaves accessible from $q$ are $\{1,3,4,7,14,15\}$, and we have $A(q)=\{1,3,4,7\}$. 
%An accessible sample for the node containing $q$ is $\{3,11,12,14,15\}$, one may check that the label of this node is isomorphic to the subgraph of  $\G(T,\mathcal{F})$ induced by this accessible sample.
}
\label{fig:GLTexample}
\end{figure}

\begin{definition} [\cite{GP07, GP08}]
A \emph{graph-labelled tree} (GLT) is a pair $(T,\mathcal{F})$, where $T$ is a tree and $\mathcal{F}$ a set of graphs, such that each node $u$ of $T$ is \emph{labelled} by the graph $G(u) \in \mathcal{F}$, and there exists a bijection $\rho_u$ between the edges of $T$ incident to $u$ and the vertices of $G(u)$.  (See Figure~\ref{fig:GLTexample}.)
\end{definition}

When we refer to a node $u$ in a GLT $(T,\mathcal{F})$, we usually mean the node itself%
%Marc-v6
, although we may sometimes use the notation $u$ as a shorthand for its label $G(u) \in \mathcal{F}$, the meaning being clear from context; for instance, notation will be simplified by saying $V(u) = V(G(u))$. 
The vertices in $V(u)$ are called \emph{marker vertices}, and the edges between them in $G(u)$ are called \emph{label-edges}. 
%Marc-v6: I find the terminology "vertices" for the endpoints of label edges confusing.  Can we just say "marker vertices"?
%Eme-v7: you are right but when considering just a graph G=G(u) it will be strange to say "the marker vertices q and r of the edge e of G" since marker vertex is defined for GLTs not for a single graph... so I just added "(marker)" below
For a label-edge $e=uv$ we may say that $u$ and $v$ are \emph{the (marker) vertices of} $e$. 
% MT 08/28/12 - Reviewer pointed out that tree-edges are already defined.
%The edges of $T$ are \emph{tree-edges}.   
%
%Marc-v6: Introduced the definition before knowing what it relates to.  Common sentence construction for French translations, but doesn't work as well in English.  See my note later on this for more.  Recurs throughout the document.
For the internal tree-edge $e = uv$, we say the marker vertices $\rho_u(e)$ and $\rho_v(e)$ are the \emph{extremities} of $e$.  
For convenience, we may say that a tree-edge and its extremities are \emph{incident}. 
Furthermore, $\rho_v(e)$ is the \emph{opposite} of $\rho_u(e)$ (and vice versa).  A leaf is also considered an \emph{extremity} of its incident tree-edge, and its opposite is the other extremity of that tree-edge (marker vertex or leaf). 
% MT 08/31/12 - Reviewer noted that we sometimes use "opposite" without the article, e.g. "if q is the marker vertex opposite a leaf l...".  It's a subtle English thing that is technically correct, and I prefer it because it streamlines sentences.  But I understand the confusion to a non-native speaker so hear I introduce this usage.
Sometimes a marker vertex will simply be said to be \emph{opposite} a leaf or another marker vertex, the  meaning in this case being that implied above.
If $q$ is a marker vertex 
%Marc-v6: In the definition below you use T(q), but you haven't defined it yet.
%DGC - added a reference to Figure 1.
such that $\rho_u(e)=q$, then we let $L(q)$ denote the set of leaves of the tree not containing $u$ in the forest $T-e$; see Figure~\ref{fig:GLTexample}
where $L(q) = \{1,2,3,4,5,6,7\}$.  
% MT 08/28/12 - Reviewer noted subtle errors if definitions not extended for leaves.
%\xtofComment{Concerning the extension of these notions to leaves, I would prefer to follow referee 1's suggestion (that is modifying GLT's def: leaves are labeled by the one-vertex graph and can be considered as marker vertices - see below also}
Extending this notion to leaves, the set $L(\ell)$ for the leaf $\ell$ is equal to all leaves in $T$ different from $\ell$.
% MT 08/28/12 - Reviewer noted only use T(q) for following definition.  Removing T(q) and updating definition for simplicity.
%The subtree of $T$ spanning the leaves in $L(q)$ is denoted $T(q)$.
%
The central notion for GLTs with respect to split decomposition is that of \emph{accessibility}:

%\xtofComment{I don't really like the definition below as it is. I think the purpose of introducing $T(q)$ was to constrain in the definition the first two marker vertices to not belong to the same node. Formally it may not be important. But still observe the confusion of the referee with respect to Figure 1. And this is how we will used the accessibility definition. So I prefer to use referee's suggestion as a definition, which would give :\\
%\emph{Two marker vertices $q$ and $q'$ are \emph{accessible} from one another if there is a sequence $\Pi$ of marker vertices $q_0,\ldots,q_{2k+1}$ of marker vertices such that: 1) $q_0=q$ and $q'=q_{2k+1}$; 2) $q_{2i}$ and $q_{2i+1}$ are the two extremities of a tree edge($0\leqslant i\leqslant k$); and 3) $q_{2i+1}q_{2i+1}$ is a label edge.}\\
%Also I double checked the non-use of $T(q)$ later in the paper.\\
%Finally, Marc's extension to leaves was consistent with its modification, but not with the one I propose. This is one the reasons I really like the idea of defining/considering a leaf of a GLT as both a leaf and a marker vertex (i.e. a leaf node is labeled by the trivial one vertex graph. By the way, this is often how I explained the structure in presentation. But this may have impact on the split paper. Still I think it deserves the effort.
%}

\begin{definition} [\cite{GP07, GP08}]
Let $(T,\mathcal{F})$ be a GLT.  Two marker vertices $q$ and $q'$ 
%Marc-v6: I don't think the notation T(q) has been defined yet.  Fixed above.
% MT 08/28/12 - Reviewer notes only use T(q) here and I don't think it's needed for the definition so am getting rid of it.  Also noted that adding it to this definition contradicted the caption in Figure~1, as noted by the reviewer.
%such that $q\in T(q')$ and $q'\in T(q)$
%
%(possibly $q = q'$) 
are \emph{accessible} from one another if there is a sequence $\Pi$ of marker vertices $q,\ldots,q'$ such that:

\begin{enumerate}
\item every two consecutive elements of $\Pi$ are either the vertices of a label-edge or the extremities of a tree-edge;
\item the edges thus defined alternate between tree-edges and label-edges.
\end{enumerate}

% MT 08/28/12 - Reviewer questioned how "naturally" it extended.  Now making it explicit.
%DGC - added a reference to Figure 1.
Two leaves are accessible from one another if their opposite marker vertices are accessible; 
similarly for a leaf and marker vertex being accessible from one another; see Figure~\ref{fig:GLTexample} where the leaves accessible from
$q$ include both 3 and 15 but neither 2 nor 11.
%This definition naturally extends when the first and/or the last element of $\Pi$ is a leaf (providing \emph{accessibility} between two leaves, or between a leaf and a marker vertex). 
By convention, a leaf or marker vertex is accessible from itself.
\end{definition}

Note that, obviously, if two leaves or marker vertices are accessible from one another, then the sequence $\Pi$ with the required properties is unique, and the set of tree-edges in $\Pi$ forms a path in the tree $T$.
%
%Marc-v6: Anytime a sentence "changes direction" like this, you have to put a comma.  This was a common mistake throughout the paper and I will only comment on it here.
%DGC slight rewording and added a reference to Figure 1.
If $q$ is a marker vertex, then we let $A(q)$ denote the set of leaves in $L(q)$ accessible from $q$; see Figure~\ref{fig:GLTexample}.  
% MT 08/28/12 - Reviewer noted subtle errors in statements if definition not extended to leaves.
The set $A(\ell)$ is similarly defined for a leaf $\ell$.

\begin{definition} [\cite{GP07, GP08}]\label{def:accessibility-graph}
Let $(T,\mathcal{F})$ be a GLT.  Then its \emph{accessibility graph}, denoted $\G(T,\mathcal{F})$, is the graph whose vertices are the leaves of $T$, with an edge between two distinct vertices  if and only if the corresponding leaves are accessible from one another. Conversely, we may say that $(T,\mathcal{F})$ \emph{is a GLT of} $\G(T,\mathcal{F})$.
\end{definition}

Accessibility allows us to view GLTs as encoding graphs; an example appears in Figure~\ref{fig:GLTexample}.  The following remarks directly follow from Definition~\ref{def:accessibility-graph}:

%Marc-v6: why not "if and only if"?
%Eme-v7: right, it is better with "if and only if"
\begin{remark} \label{connectedLabels}
%If a graph $G$ is connected, then every label in a GLT of $G$ is connected.
A graph $G$ is connected if and only if every label in a GLT of $G$ is connected.
\end{remark}   

\begin{remark}\label{existsAccessible}
Let $(T,\mathcal{F})$ be  a GLT, with $\G(T,\mathcal{F})$ connected. 
%Marc-v6
%DGC removed "we have that"
For every marker vertex $q$ in $(T,\mathcal{F})$, $A(q)$ is non-empty.
%
%there exists a leaf in $L(q)$ accessible from $q$.
\end{remark}  

\begin{remark}\label{rem:accessible-split}
%Marc-v6
Let $e$ be an internal 
tree-edge of a GLT $(T,\mathcal{F})$, with $\G(T,\mathcal{F})$ connected, and let $p$ and $q$ be the two extremities of $e$. Then  the bipartition $(L(p),L(q))$ is a split of $\G(T,\mathcal{F})$. Moreover $A(q)$ and $A(p)$ are the frontiers of that split.
\end{remark}
%DGC the first L(q) above changed to L?.

\begin{remark}\label{subgraph}
Let $(T,\mathcal{F})$ be a GLT, with $\G(T,\mathcal{F})$ connected. For every graph label $G(u)$ in $\mathcal{F}$, there exists a subset $L$ of leaves of $T$ such that $G(u)$ is isomorphic to the subgraph of $\G(T,\mathcal{F})$ induced by $L$. Note that $L$ can be built by choosing, for every vertex $q$ of $G(u)$, an element of $A(q)$.
\end{remark}

%\com{*** accessible samples defined below are now useless, I would remove them...   ***}
%\com{***  they are mentioned in Section 5 but always as an alternative that can be omitted ***}

%\begin{definition}
%Let $u$ be a node in a GLT.  An \emph{accessible sample} for $u$ is a set of leaves $L$~\mbox{such that}
%for every marker vertex $q$ of $u$, we have $L(q)\cap L=\{e\}$ where $e$ is accessible from $q$.
%%$L(q)\cap L$ has only one element, and this element is accessible from $q$.
%%\begin{enumerate}
%%
%%\item for every $q \in V(u)$, there is a unique $\ell_q \in A(q)$ such that $\ell_q \in L$;
%%
%%\item $|V(u)| = |L|$.
%%
%%\end{enumerate}
%\end{definition}

%\com{*** degenerate split/join are never used (the minimality of the split-tree suffices to prove Prop \ref{circlePartition}) ***}
%\com{*** also, the formal definition of node-split is useless  ***}

%\begin{definition} \label{def:node-split}
%The \emph{node-split} 
%%%%%%%(for ``node-split'') 
%is the inverse of the node-join.  More precisely, let $v$ be a node such that $G(v)$ contains the split $(A,B)$ with frontiers $A'$ and $B'$.  The node-split with respect to $(A,B)$ replaces $v$ with two new adjacent nodes $u$ and $u'$ labelled by $G[A] + q$ and $G[B] + q'$, respectively, where $q$ and $q'$ are the extremities of the new tree edge thus created, $q$ being universal to $A'$, and $q'$ being universal to $B'$. The extremities of the tree edges incident to $v$ remain unchanged. 
%See Figure~\ref{fig:nJoinNsplitExample}.
%\end{definition}
%\com{*** I have voluntarily skipped the formal def of node-split ***}

\begin{figure}[h]  %[htbh]
\begin{center}
\includegraphics[scale=0.70]{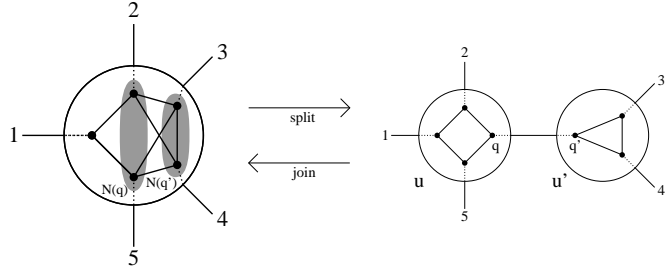}
\end{center}
\vspace{-0.5cm}
\caption{Example of the node-join and node-split.}
\label{fig:nJoinNsplitExample}
\end{figure}

Let $e=uu'$ be an internal tree-edge of a GLT $(T,\mathcal{F})$, and let $q \in V(u)$ and $q' \in V(u')$ be the extremities of $e$.  The \emph{node-join} of 
%Marc-v6: do you really mean a comma here?  I guess it's the difference of defining it with respect to pairs of nodes versus edges.  Maybe edges would be more aesthetically pleasing?
%Eme-v7: I think we specify nodes for this operation, so I replace "u,u'" with "u and u'"
%$u,u'$
$u$ and $u'$
is the following operation: contract the tree-edge $e$, yielding a new node $v$ labelled by the join between $G(u)$ and $G(u')$ with respect to $q$ and $q'$. Every other tree-edge and 
%Marc-v6
their pairs of extremities are preserved. 
The \emph{node-split} is the inverse of the node-join. Both operations are illustrated 
%Marc-v6: Recurring problem
%on
in
Figure~\ref{fig:nJoinNsplitExample}. A key property to observe is that the node-join operation and the node-split operation preserve the accessibility graph of the GLT.

To end this subsection, we recall the main result of split decomposition theory \cite{Cun82}, 
%stated as in \cite{GP07}\cite{GP08} in the GLT setting:  
%Marc-v6: A little cumbersome above.
which we restate below in terms of GLTs, as in~\cite{GP07, GP08}:

\begin{theorem} [\cite{Cun82, GP07, GP08}]
\label{thm:splitTheorem}
For any connected graph $G$, there exists a unique graph-labelled tree $(T,\mathcal{F})$ whose labels are either prime or degenerate, having a minimal number of nodes, and such that $\G(T,\mathcal{F})=G$. 
%Marc-v6: I think it's bad practice to have definitions in theorems.  I know that Derek feels this way.  See new definition below.
%Eme-v7: I agree and feel this way too
%We call \emph{split-tree} of $G$ this GLT and it is denoted $ST(G)$. 
\end{theorem}

%Marc-v6: see comment above in previous lemma.
\begin{definition}
The unique graph-labelled tree guaranteed by Theorem~\ref{thm:splitTheorem} is called the \emph{split-tree} for $G$, and is denoted $ST(G)$.
\end{definition}

For example, the GLT in Figure~\ref{fig:GLTexample} is the split-tree for the accessibility graph pictured there. The split-tree of a graph $G$ could be thought as a representation of the set of splits: it is known that every split either corresponds to a tree-edge of the split-tree or to the tree-edge resulting from a node-split of some degenerate-node (for more details, the reader should refer to the companion paper~\cite{GPTC11a}).

%----------------------------------------------------------------------------------------------------------------------
\subsection{Lexicographic Breadth-First Search}
\label{sub:prel3}

Lexicographic Breadth-First Search (LBFS) was developed by Rose, Tarjan, and Lueker for the recognition of chordal graphs~\cite{RTL76}, and has since become a standard tool in algorithmic graph theory~\cite{C04}.  It appears here as Algorithm~\ref{alg:LBFS}.

%An \emph{ordering} $\sigma$ of a graph $G$ is a linear ordering of its set of vertices $V(G)$.
%Formally, we can define it either as an injective mapping from $V(G)$ to the integers, or as an ordering binary relation. We slightly abuse notation by allowing $\sigma$ to represent such a mapping as well as the ordering, and we let $<_\sigma$ denote the binary relation: $x<_{\sigma} y$ is equivalent to $\sigma(x)<\sigma(y)$.
%In such a case, we say that ``$x$ appears before $y$'', or ``earlier than $y$'', in $\sigma$. Similarly,  by ``first'', ``last'' and ``penultimate'', we denote respectively,  the smallest element of $<_\sigma$, the greatest element and the element appearing immediately before the last one. 
%

% By an LBFS ordering of the graph $G$, we mean any ordering produced by Algorithm~\ref{alg:LBFS} on input  graph $G$. 

\begin{algorithm} 
\KwIn{A graph $G$ with $n$ vertices.}
\KwOut{An ordering $\sigma$ of $V(G)$ defined by a mapping $\sigma : V(G) \rightarrow \{1,\ldots,n\}$.}
%\KwOut{A numbering of $V(G)$ inducing the ordering $\sigma$.}

\BlankLine
\lForEach{$x \in V(G)$} {label($x$) $\gets \epsilon$ (the empty-string) \;}
\SetKw{DownTo}{downto}
\For{$i = 1$ \KwTo $n$} {
	pick an unnumbered vertex $x$ with lexicographically largest label\;
	$\sigma(x) \gets i$ \tcp*[l]{assign $x$ the number $i$} 
	\lForEach{unnumbered vertex $y \in N(x)$} {append $n - i + 1$ to label($y$)\;}
	}

\caption{Lexicographic Breadth-First Search} \label{alg:LBFS}
\end{algorithm}

% MT 08/31/12 - Reviewer didn't like the "definition of LBFS" for a few reasons.  First, they didn't like defining LBFS via an algorithm rather than some formal linear properties.  They also said the algorithm was unclear because the selection of a vertex at each stage is non-deterministic -- i.e. we don't require a unique vertex with largest label.  They wanted to know why different LBFS orderings were possible for a single graph.  Regarding the first point, since LBFS is an algorithm, I don't see how we could define it by anything but an algorithm.  What the reviewer really wants is a definition of LBFS orderings.  Or rather, since we provide one, they want a characterization of these orderings.  Although we don't use the 4-vertex characterization in this paper, it will likely clarify some of the perceived "imprecision" relating to LBFS orderings.  For the sake of readers unfamiliar with LBFS, It is also worth taking up the reviewer's suggestion to discuss its non-determinism with examples.  The following paragraph has been rewritten to accommodate these concerns.
%DGC I've made a few minor changes to Marc's paragraph:  "empty string" rather than "null" and added a "the".  I've also added a reference to Golumbic in the following lemma.  DNB proved it in one direction, Golumbic in the other.  I've also changed Marc's last sentence - in particular I didn't like the "Despite this.."
By an LBFS ordering of the graph $G$ (or its set of vertices $V(G)$), we mean any ordering $\sigma$ produced by Algorithm~\ref{alg:LBFS} when the input is $G$. We write $x<_{\sigma} y$ if $\sigma(x)<\sigma(y)$.  Notice that the first vertex in any LBFS ordering is arbitrary.  This is because all vertices start out with the empty string label.  More generally, the vertex with the lexicographically largest label may not be unique.  As another example, if $x$ is numbered first, meaning it is the first vertex in the LBFS ordering, then every vertex in $N(x)$ will share the lexicographically largest label at the time the second vertex is  numbered.  In other words, any vertex in $N(x)$ can follow $x$ in an LBFS ordering.  Interestingly, LBFS orderings can be characterized as follows:  

\begin{lemma} [\cite{DNB96}\cite{Gol04}]\label{4vertex}
An ordering $\sigma$ of a graph $G$ is an LBFS ordering if and only if for any triple of vertices $a <_{\sigma} b <_{\sigma} c$ with $ac \in E(G)$, $ab \notin E(G)$, there is a vertex $d <_{\sigma} a$ such that $db \in E(G)$, $dc \notin E(G)$.
\end{lemma}  

%By an LBFS ordering of the graph $G$ (or its set of vertices $V(G)$), we mean any ordering $\sigma$ produced by Algorithm~\ref{alg:LBFS} when the input is $G$. We write $x<_{\sigma} y$ if $\sigma(x)<\sigma(y)$. 

For a subset $S$ of $V(G)$, we denote $\sigma[S]$ 
%Marc-v6
as the restriction of $\sigma$ to $S$: that is, for $x, y\in S$, $x<_{\sigma[S]} y$ if and only if $x<_{\sigma} y$. 
A \emph{prefix} of $\sigma$ is a subset $S$ such that $x<_{\sigma}y$ and $y\in S$ implies that $x\in S$.

The following remarks are obvious and well-known observations:

\begin{remark} \label{universalStillLBFS}
If $\sigma$ is an LBFS ordering of a graph $G$, and $x$ is a universal vertex in $G$, then $\sigma[V(G) - \{x\}]$ is an LBFS ordering of $G - x$.
\end{remark}

\begin{remark} \label{prefix}
Let $S$ be a prefix of any LBFS ordering $\sigma$ of a graph $G$.  Then $\sigma[S]$ is an LBFS ordering of $G[S]$.
\end{remark}

Our circle graph recognition algorithm is based on special properties of \emph{good vertices} and on the hereditary property of LBFS orderings with respect to the label graphs of a GLT (and thus of the split-tree).
%xtof-v8: completed sentence above

\begin{definition}
A vertex $x \in V(G)$ is \emph{good} for the graph $G$ if there is an LBFS ordering of $G$ in which $x$ appears last.
\end{definition}

%Marc-v6: See comment below in the following lemma.
%xtof-v8: I made the statement consistent with the split paper
\begin{definition} [Definition 3.5 in \cite{GPTC11a}]
Let $u$ be a node of a GLT $(T,\mathcal{F})$ and let $\sigma$ be an LBFS ordering of $G=Gr(T,\mathcal{F})$. For any marker vertex $p$, let 
\emph{$x_p$} be the earliest vertex of $A(p)$ in $\sigma$. Define \emph{$\sigma_u$} to be the ordering of $G(u)$ such that for $q,r\in V(u)$, $q<_{\sigma_u} r$ if $x_q<_{\sigma} x_r$.
%Let $\sigma$ be an LBFS ordering of $G$, and let $u$ be a node in $ST(G)$.  Then we define $\sigma_u$ to be the vertex ordering of $V(u)$ such that for any pair $q,q' \in V(u)$:
%\vspace{-0.2cm}
%$$q<_{\sigma_u} q' \Leftrightarrow \exists x\in A(q), \exists x'\in A(q') \mbox{ such that } x<_{\sigma} x'.$$
\end{definition}
%

% XTOF - VERIFIER DANS LA PARTIE ALGO SI C'EST UTILE
%The next result come from the companion paper \cite{}.

\begin{lemma} [Lemma 3.6 in \cite{GPTC11a}] \label{inducedLBFS}
%xtof-v8: I made the statement consistent with the split paper
Let $\sigma$ be an LBFS ordering of graph $G=Gr(T,\mathcal{F})$, and let $u$ be a node in $(T,\mathcal{F})$. Then $\sigma_u$ is an LBFS ordering of $G(u)$.
%Let $\sigma$ be an LBFS ordering of $G$, and let $u$ be  a node in $ST(G)$; 
% Marc-v6: Definition in lemma problem again, see new definition above.
%If $\sigma_u$ is the vertex ordering of $V(u)$ such that for any $q,q'\in V(u)$:
%\vspace{-0.2cm}
%$$q<_{\sigma_u} q' \Leftrightarrow \exists x\in A(q), \exists x'\in A(q') \mbox{ such that } x<_{\sigma} x'$$
%then $\sigma_u$ is a LBFS ordering of the label graph $G(u)$.
\end{lemma}

%\com{*** I do not understand well the next corollary ***}
%\com{*** it is used in algorithm section, not sure about its use***}

%\begin{corollary}\label{cor:good}
%Let $q \in V(u+x)$ be the marker vertex opposite $x$.  Then $q$ is good for $G(u+x)$.
%\end{corollary}

%\begin{proof}
%Let $\sigma$ be an LBFS of $G+x$ in which $x$ appears last.  For every $t \in V(u+x)$, associate the leaf in $A(t)$ appearing earliest in $\sigma$, and let $L$ be the set of all such leaves.  Notice that $x \in L$.  It follows that $x$ appears last in $\sigma[L]$.  Therefore $q$ is good for $G(u+x)$, by lemma~\ref{inducedLBFS}, and the correspondence between $G[L]$ and $G(u+x)$.
%\end{proof}

%----------------------------------------------------------------------------------------------------------------------
\subsection{Circle Graphs}
\label{sub:circle}

%Marc-v6: Something here to connect the title of the subsection to what follows.
We will work with circle graphs using a variant of the double occurrence words mentioned in the introduction.
A \emph{word} over an alphabet $\Sigma$ is a sequence of letters of $\Sigma$. 
If $S$ is a word over $\Sigma$, then $S^r$ denotes the reversed sequence of letters. 
%Eme-v7 def added below, it was already implicit in what follows, but to be sure... 
The concatenation of two words $A$ and $B$ is denoted $AB$. 
% MT 08/28/12 - Reviewer didn't like circular word being defined as a sequence.
A \emph{circular word} $C$ over an alphabet $\Sigma$ is a circular sequence of letters of $\Sigma$; they can be represented by a word $S$ by considering that the first letter of $S$ follows its last letter.  That 
%A \emph{circular word} $C$ over an alphabet $\Sigma$ is represented by a sequence $S$ of letters of $\Sigma$ that can be arranged clockwise around a circle, that
%Marc-v6
% precisely 
is: if $S$ is the concatenation $AB$ of the words $A$ and $B$, then $BA$ represents the same circular 
%Marc-v6: would \cdot be better instead of the period?  I found the period jarring while I read because of its dual use to end a sentence.
%Eme-v7: ok let us just remove the dot
word $C$ 
%Marc-v6
as $S=AB$, and we denote this by $C\sim AB\sim BA$. 
%
%and we may denote $AB= BA$ in terms of circular words. 
%A \emph{factor} of \xtofModif{ an empty} word, 
A \emph{factor} of a word
%Marc-v6: I only remember seeing "resp" in parentheses.  It helps by being cleaner to read.
%Eme-v7: as you wish but I have also often read it between commas
(respectively of a circular word), 
over $\Sigma$ is a sequence of consecutive letters in this word
%Marc-v6: as above
(respectively in a word representing this circular word).
%Eme-v7 precision added below, not much useful, but it is true that we make this abuse sometimes
%%%TO-BE-DONE please check my english below
Formally, we may sometimes make the abuse to consider a factor of a given (circular) word as a set of letters, and conversely, as soon as this set of letters forms a factor in  this (circular) word.
If the sequence $S$ of elements of $\Sigma$ defines a circular word $C$, then the reversed sequence $S^r$ defines the \emph{reflection} of $C$, denoted $C^r$.  

\begin{figure}[h] %[htbh]
\begin{center}
%\scalebox{0.5}{\input{circleExample.pstex_t}}
\includegraphics[scale=1]{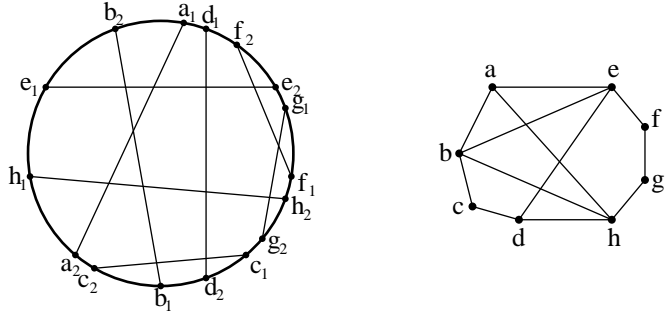}
\end{center}
%\vspace{-0.2in}
\vspace{-0.5cm}
\caption{A  chord diagram $C$ drawn on a circle (on the left) and the corresponding circle graph $G$ (on the right). By convention we read the sequences clockwise from figures. We have $C(a_2,f_2)=h_1e_1b_2a_1d_1$ and $C(f_2,a_2)=e_2g_1f_1h_2g_2c_1d_2b_1c_2$.}
\label{fig:circleGraphExample}
\end{figure}

%Marc-v6: You've already introduced the notion of chord diagram in the introduction, plus, the geographic notion may already be familiar to those reading the paper.  Since you are therefore "redefining" them here, I think it's important to help the reader understand this.
%Eme-v7: usually the introduction is not thought of as being really part of the paper, the terms used there can be defined properly later with possible variants and precisions... a def in the intro is just the "idea" behind the forthcoming formal def. I just change here the beginning and end of your sentence. I also changed the beginning of the intro to avoid such confusion
We define formally 
%We redefine 
the chord diagrams mentioned in the introduction using circular words.
%; doing so will simplify the proofs later in the paper.
%
For a set $V$, called
%Marc-v6
a 
set of \emph{chords}, a \emph{chord diagram} on $V$ is a circular word on the alphabet $\V=\bigcup_{v\in V}\{v_1,v_2\}$ where every letter appears exactly once. The elements of $\V$ are called \emph{endpoints}, and, for every chord $v\in V$, the letters $v_1$ and $v_2$ of $\V$ are called \emph{the endpoints of} $v$.  
% MT 08/28/12 - Agreed with reviewer that geometric interpretation should be moved here to reinforce concepts.
Geometrically, a chord diagram can be represented as a circle inscribed by a set of chords (see figure~\ref{fig:circleGraphExample}).  
%Marc-v6: needed for later in the text.  
%Eme-v7: "for G" not defined yet, I put it later
%Notice that if $C$ is a chord diagram for $G$, then $C^r$ is a chord diagram for $G$ as well.  
Now,
if $C$ is a chord diagram on $V$, then the \emph{simple} chord diagram induced by $C$ is the circular word $\bar C$ on $V$ obtained by replacing the endpoints appearing in $C$ by the corresponding chords of $V$ 
%Marc-v6
(or equivalently, removing the subscripts from the endpoints).
%Eme-v7 precision below useless, and the figure shows this equivalence (and the subscripts are just a formal trick). To link those defs I put again a sentence that you deleted
% MT 08/28/12 - Moved sentence earlier as requested by reviewer.
%Geometrically, a (simple) chord diagram can be represented as a circle inscribed by a set of chords.
% -- here we see the equivalence with the chord diagrams from the introduction.
%
  If $a$ and $b$ are two endpoints of the chord diagram $C\sim AaBbA'$, with $A,B,A'$ words on $\V$, then we define the factor $C(a,b)=B$. 
% MT 08/28/12 - Reviewer thought it better to emphasize this is a result of the just defined notion of factor, not a separate notion on its own.
Based on this, it follows that $C(b,a) = A'A$, and similarly
% and $C(b,a)=A'.A$
%Marc-v6
% on $\V$
%, similarly 
$C^r(a,b)=A^rA'^r$ and $C^r(b,a)=B^r$.

%We will also use the following notations:
%\com{*** to be harmonized along the paper: replace marc's arrow notation (where a subscript for the diagram is missing) with mine more standard and concatenable, then delete this alternative notation ***}
%
%$$\W(C;a,b)=a \xrightarrow{c} b$$
%$$\W(C;b,a)=a \xrightarrow{cc} b$$

% MT 08/28/12 - Reviewer wanted distinction drawn between vertices and chords.
The chord diagram $C$ encodes the graph $G = (V,E)$ as follows: the chords of $C$ correspond to the vertices $V$, two of which are adjacent if and only if their corresponding chords intersect.  Using the notation from above, vertices $x$ and $y$ are adjacent if and only if
%The chord diagram $C$ on $V$ encodes the graph $G=(V,E)$ as follows: two vertices $x$ and $y$ of $V$ are adjacent if and only if the chords $x$ and $y$ \emph{intersect}, meaning that 
the factor $C(x_1,x_2)$ contains either $y_1$ or $y_2$ but not both.  The \emph{circle graphs} are the graphs that can be encoded by chord diagrams in this way.  We say that $C$ is a \emph{chord diagram for} $G$, or that $C$ \emph{encodes} $G$. The above definitions are naturally extended to simple chord diagrams. 
%Eme-v7 Marc's sentence put here
Notice that if $C$ is a chord diagram for $G$, then $C^r$ is a chord diagram for $G$ as well.
An example appears in Figure~\ref{fig:circleGraphExample}. 

% MT 08/28/12 - Reviewer again wanted distinction between vertices and chords.
If $L \subseteq V(G)$, 
%If $L$ is a subset of chords of  $C$, 
then $C[L]$ is the chord diagram formed by removing from $C$ all 
%endpoints of 
chords corresponding to vertices not in $L$.
%not in $L$.  The following is obvious:

% MT 08/28/12 - Same as above, distinction between vertices and chords.
\begin{remark} \label{inducedChord}
If $C$ is a chord diagram for $G$, and 
$L \subseteq V(G)$,
%$L$ a set of its chords
then $C[L]$ is a chord diagram for $G[L]$.
\end{remark}

\begin{figure}[h] %[htbh]
\begin{center}
\includegraphics[scale=1.4]{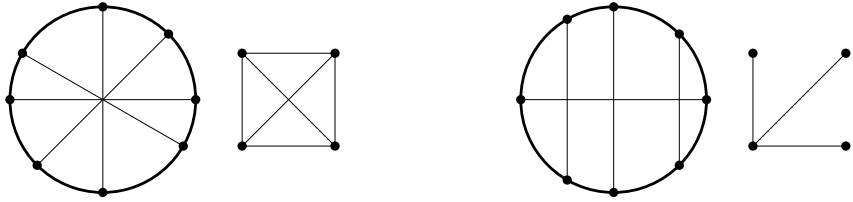}
\end{center}
\vspace{-0.5cm}
%\vspace{-0.2in}
\caption{A simple chord diagram for a clique; a simple chord diagram for a star.}
\label{fig:degenerateExample}
\end{figure}

Simple chord diagrams are in general not uniquely determined by the graph they encode, as demonstrated by the example of cliques and stars (depicted 
%Marc-v6
in 
Figure~\ref{fig:degenerateExample}). A chord diagram $C$ of a clique 
% MT 08/28/12 - Reviewer didn't think additional notation of G, \pi, c, etc. added value.  Updated throughout this paragraph.
$G$
%$G=(V,E)$ 
is of the form 
%Marc-v6
$C\sim AA$,
where 
$A$
%$A=\pi(V)$ 
%DGC  "of" added - and period at the end of the sentence.
is any permutation of its vertices.
% of $V$. 
If $G$ is a star with centre vertex $c$, a chord diagram $C$ is of the form 
%Marc-v6: Please check these corrections to make sure that I am right.
$C\sim c A c  A^r$, where 
$A$
%$A=\pi(V\setminus\{c\})$ 
is any permutation of 
the non-centre vertices of the star.
%$V\setminus\{c\}$ and  $A^r$ its reversal.
%
%In the former, the chords must pairwise intersect; in the latter, there must be one chord (the centre) that intersects all others, none of which intersect each other. 
In both cases, one can  transpose any two chords (distinct from the centre, in the case of a star). 
% EG 09/23/12 at least five vertices, or not ?
%XXX
%On the contrary, it is known that a circle graph is prime (i.e. has no split) if and only if it has a unique simple chord diagram (up to reflection)~\cite{Bou87} (see also \cite{Cou08}).
On the contrary, it is known that if a circle graph is prime (i.e. has no split) 
 then it has a unique simple chord diagram (up to reflection),
 %DGC changed ">5" to ">4" below.
 and that the converse is true provided the graph has more than four vertices~\cite{Bou87} (see also \cite{Cou08}).

%Marc-v6: My taste is for symmetry in the diagrams.  Right now the chords don't look great.  But I really don't care about this.
%Eme-v7 not sure to understand this remark : how can you have a symmetry  for chords ? I guess you mean regular distances between the endpoints, I agree but this figure was drawn to be compatible with figure 5, where you obtain figure 4 just by putting two circle graphs one on the other. You lose regularity, but win a direct illustration of the circle join without modifying distances... 

%This section is almost a preliminary one. It presents the key concepts for our further constructions. In particular all our technique rely on the next definition.
The concept of join between 
%Marc-v6
two 
graphs $G$ and $G'$ with respect to two vertices $q\in V(G)$ and $q'\in V(G')$ 
%Marc-v6
was
 defined in Section~\ref{sub:prel1}. A similar join operation directly applies to chord diagrams. It will be thoroughly used in our incremental split-tree construction of circle graphs.

% MT 08/28/12 - Reviewer noted that the \odot and \hat\odot operations are symmetric.  They wondered if it really added to the understanding of the paper.  On the contrary, they noted that it was yet one more thing to remember in an already technical paper.  I happen to agree strongly with this.  The few times I have read this parameterization of the algorithm (remember, always out of context while I was working elsewhere), I have constantly had to return to the definition to remind me of its meaning, whereas \odot is intuitive and easy to remember.  Now, making such a change will require significant (albeit mostly typographical) changes to the paper, and may be contentious with others.  I'll leave it in for now and merely state my opinion in favour of eliminating the \hat\odot operator.
\begin{definition}\label{def:circle-join}
% XTOF 07/09/2012 : I agrree with reviewer 1 suggestion on how to define the second circle-join.
%Let $C$ and $C'$ be chord diagrams on $V$ and $V'$, respectively, and let $q\in V$ and $q'\in V$. 
%We define two \emph{circle-join} operations between $C$ and $C'$ w.r.t. $q$ and $q'$ as follows
%$$(C,q)\odot (C',q')\ \sim \ C(q_1,q_2)C'(q'_1,q'_2)C(q_2,q_1)C'(q'_2,q'_1)$$
%$$(C,q)\ \hat\odot\  (C',q')\ \sim \ C(q_1,q_2)C'(q'_2,q'_1)C(q_2,q_1)C'(q'_1,q'_2)$$
%DGC  Two "belongs to" below - and more importantly the first sentence ends with V' not V.
Let $C$ and $C'$ be chord diagrams on $V$ and $V'$, respectively, and let $q$ belong to $V$ and $q'$ belong to  $V'$. 
We define a \emph{circle-join} operation between $C$ and $C'$ with respect to $q$ and $q'$ as follows
$$(C,q)\odot (C',q')\ \sim \ C(q_1,q_2)C'(q'_1,q'_2)C(q_2,q_1)C'(q'_2,q'_1)$$

\end{definition}

\begin{figure}[h]
\begin{center}
%\scalebox{0.5}{\input{cJoinExample.pstex_t}}
\includegraphics[scale=0.85]{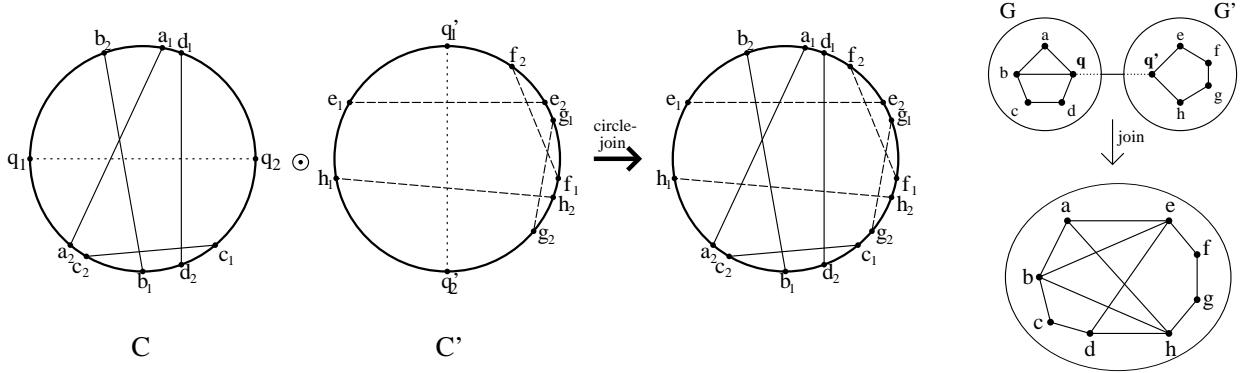}
\end{center}
\vspace{-0.5cm}
\caption{A circle-join $(C,q)\odot (C',q')$, and the corresponding node-join between graphs $G$ and $G'$ encoded by $C$ and $C'$ respectively (Remark \ref{c-join}).}
\label{fig:cJoinExample}
\end{figure}

%\xtofModif{
Observe that the circle-join is not commutative. We may use the notation $(C,q)\ \hat\odot\  (C',q')$ instead of $(C',q')\ \odot\ (C,q)$.
%}
%%%%REPRENDRE
%Observe that, by construction, the resulting sequences of letters define chord diagrams on the set of chords $(V\setminus\{q\})\cup(V'\setminus\{q'\})$. 
By construction, the resulting sequences of letters define chord diagrams on the set of chords $(V\setminus\{q\})\cup(V'\setminus\{q'\})$. 
An illustration of this construction and of the obvious 
%Marc-v6: agree it is obvious enough not to deserve a lemma/citation.  I have made it a "remark" below and fixed the future references.
remark below
% 
%\cite{easy} 
 is given in Figure~\ref{fig:cJoinExample}. 
%
%An illustration of this construction and of the well-known Lemma \ref{c-join} below is given Figure~\ref{fig:cJoinExample}. 

%Lemme 8 de \cite{Cou08}
\begin{remark} \label{c-join} %\cite{easy} %\cite{Cou08,Spi}
Let $C$ and $C'$ be chord diagrams of $G$ and $G'$, respectively, and let $q$ belong to $V(G)$ and $q$ belong to  $V(G')$. 
The chord diagrams $(C,q)\odot (C',q')$ and $(C,q)\ \hat \odot\ (C',q')$ encode the graph $H=(G,q)\otimes (G',q')$. %resulting from the join between $G$ and $G'$ w.r.t. $q$ and $q'$.
\end{remark}
Assume that $G'$ and $C'$ are as in Remark~\ref{c-join}.
Let us also remark that, as $C'^r$ is a chord diagram of $G'$, 
%Marc-v6
both
 $(C,q)\odot (C'^r,q')$ and $(C,q)\ \hat \odot\ (C'^r,q')$ are also chord diagrams of the same graph $H$.  
%Marc-v6: Moved the result below from the end of section 3.1 and reworked it with a proof to help the reader.  I didn't think this was so trivial as to avoid citation/proof.
%Eme-v7 ok 
Finally, Remark~\ref{c-join} also allows us to obtain the following well-known result, restated in terms of graph-labelled trees:

\begin{corollary} 
\label{cor:circle-labels}
% MT 08/28/12 - Reviewer noted that because the definition of GLTs doesn't require a correspondence between F and nodes in T, the corollary is technically false.  Didn't want to unnecessarily burden an already technical GLT definition, so I tightened the statement of the corollary.
Let $(T,\mathcal{F})$ be a GLT. The accessiblity graph $\G(T,\mathcal{F})$ is a circle graph if and only if for every node $u$ in $T$, the label $G(u)$ is a circle graph.
%every label graph of $\mathcal{F}$ is a circle graph.
\end{corollary} 

\begin{proof}
Notice that by recursively performing node-joins, any GLT can be reduced to a single node labelled by its accessibility graph.  Thus, by Remark~\ref{c-join}, if every label in the GLT is a circle graph, then so is its accessibility graph.  For the converse, notice that every label in a GLT is isomorphic to an induced subgraph of its accessibility graph (Remark~\ref{subgraph}).  Since every induced subgraph of a circle graph is also a circle graph (Remark~\ref{inducedChord}), if $G$ is a circle graph, so is every label in a GLT having $G$ as its accessibility graph.
\end{proof}
%

%----------------------------------------------------------------------------------------------------------------------
%----------------------------------------------------------------------------------------------------------------------
%\section{Consecutiveness, chord diagrams and LBFS}
\section{Consecutiveness and LBFS Incremental Characterization}
\label{sec:consecutivity}

The key technical concept for this paper is given by the definition below of consecutiveness.
%Marc-v6
%Next, 
%DGC Removed "The"
Subsections \ref{sec:circle-join-preserving}, \ref{sec:circle+split} and \ref{sec:good}
that follow are independent from each other and provide general properties of circle graphs with respect to consecutiveness. Their results will be merged  in Subsection \ref{sec:charact} with the incremental construction of the split-tree from \cite{GPTC11a} to get the main theorem of this section.

\begin{definition}
Let $C$ be a chord diagram on a set $V$ of chords. If a set of endpoints $S_e\subseteq \V$ is a factor of $C$ (i.e. appears consecutively), then the first
%Marc-v6
and
last endpoint in this factor 
%Marc-v6
are called \emph{bookends} for $S_e$.
\end{definition}

%Marc-v6: I don't like having two paragraphs in a definition.  It's ugly aesthetically and generally signifies two separate things that would probably benefit from having their own definitions.  I don't feel too strongly about this so change it back if you'd like to.
%Eme-v7 ok with me, it seems lighter as you did
\begin{definition}
A set of chords $S\subseteq V$ is \emph{consecutive} in $C$  if $C$ contains a set of endpoints $S_e \subseteq \V$ as a factor such that $|S_e\cap\{x_1,x_2\}|=1$ for all 
%Marc-v6
$x\in S$,
and $S_e\cap\{x_1,x_2\}=\emptyset$ for all $x\notin S$. In this case, $S_e$  \emph{certifies} the consecutiveness of $S$, and a vertex $x \in S$ is a \emph{bookend} for $S$ if one of its endpoints is a bookend for $S_e$.  The definition naturally extends to a simple chord diagram $\bar C$ by considering any chord diagram $C$ 
%Marc-v6
whose
underlying simple chord diagram is $\bar C$.
\end{definition}

%\com{*** the extension to simple chord diagrams is required for perfect formal consistency of Corollary \ref{graphToLabel}, but this is a pernicious detail ***}

Observe that if $S$ is a consecutive set of at least two chords, then two distinct chords of $S$ are bookends. On the chord diagram $C$ depicted in Figure~\ref{fig:cJoinExample}, the consecutiveness of the subset of chords $S = \{b,c,d\}$ is certified by $S_e = \{c_1,d_2,b_1\}$; the bookends of $S_e$ are $c_1$ and $b_1$, meaning $b$ and $c$ are 
%Marc-v6
bookends
for $S$.  

%----------------------------------------------------------------------------------------------------------------------
%\subsection{Circle-join Preserving consecutiveness}
\subsection{Circle-Join Property}
\label{sec:circle-join-preserving}

%Marc-v6: You can't really start a section with "Next" - it is usually only used when something has preceded it, to which a subsequent "next" will make sense.
Lemma \ref{labelToGraph} below will be crucial  
(in Section \ref{sec:charact})
%Marc-v6: little things.
for maintaining chord diagrams during the vertex insertions
%Marc-v6: I really don't like forward references.  I'm pretty sure Derek feels the same.
%Eme-v7 I feel the same when we talk about results, but here it is a forward reference on the organization of the paper, as in the introduction... Also it explains the "will" since it is disturbing to read that something 'will'  vaguely be done with no precision... and it might help the reader to know that this result will be used there, and not before. So I put it again above with '(in Section...)'
 %(see Section \ref{sec:charact})
constructing the split-tree in the companion paper~\cite{GPTC11a}. 
%Marc-v6
It is illustrated in Figure \ref{fig:reflectionExample} below. % illustrates Lemma \ref{labelToGraph}.

\begin{lemma} \label{labelToGraph}
Let $C$ and $C'$ be chord diagrams on the sets of chords $V$ and $V'$ respectively. 
Let $S\subset V$ and $S'\subset V'$ be 
sets  of chords such that $1<\mid S\mid < \mid V\mid $ and
$1<\mid S'\mid < \mid V'\mid $.
Assume that $S$ and $S'$ are consecutive in their respective chord diagrams.
%Let $S\subseteq V$ and $S'\subseteq V'$ be non-empty 
%% XTOF 07/09/2012 : as noted by reviewer, empty sets make no sense
%sets  of chords, consecutive in their respective chord 
% diagrams. 
 %
 If $q$ is a bookend of $S$, 
%
%DGC added "is" below
and $q'$ is a bookend of $S'$, then 
the set of chords $(S\setminus \{q\}) \cup (S'\setminus \{q'\})$ is consecutive in (at least) one of the following chord diagrams, 
% MT 08/28/12 - Reviewer noted that boundary cases where S and S' only contain one marker vertex aren't handled.  Updated to fix this.
with bookends being those of $S$ and $S'$ other than $q$ and $q'$:
% EG 09/23/12 I removed Marc's update at the end: , respectively (if such bookends exist).
%with bookends given by the other bookends of $S$ and $S'$ (i.e. not $q$ and $q'$):
%
$$(C,q)\odot (C',q'),~~~~~~(C,q)\ \hat\odot\ (C',q'),~~~~~~(C,q)\odot (C'^r,q'),~~~~~~(C,q)\ \hat\odot\  (C'^r,q').$$

%If $q$ is a bookend of $S$ and $q'$ a bookend of $S'$, then the set of chords $(S\setminus \{q\}) \cup (S'\setminus \{q'\})$ is consecutive (at least) one of the following chord diagrams:
%$$(C,q)\odot (C',q'),~~~~~~(C,q)\ \hat\odot\ (C',q'),~~~~~~(C,q)\odot (C'^r,q'),~~~~~~(C,q)\ \hat\odot\  (C'^r,q').$$

%Moreover, if $C(q_1,q_2)$, $C(q_2,q_1)$, and resp. $C'(q'_1,q'_2)$, $C'(q'_2,q'_1)$, contain an element which is not an endpoint of an element of $S$, resp. $S'$, then exactly one of the chord diagrams above has the required property.
\end{lemma}

\begin{proof}
%We prove the lemma assuming the last property is satisfied. If not, then the same choice as below gives a satisfying diagram, but which may not be unique.
%
% MT 08/28/12 - Reviewer noted boundary case of S and S' only containing one marker vertex is not handled. 
% EG 09/23/12 - I removed Marc's sentence below
%If $|S| = 1$ or $|S'| = 1$, then the lemma follows trivially.  So assume that $|S|,|S'| > 1$.
%
Assume that the consecutiveness of $S$ in $C$ is certified by the set $S_e$ of endpoints and, without loss of generality, let $q_1$ be the endpoint of $q$ in $S_e$. Let $r_1$ denote the other bookend of $S_e$ with chord $r$ distinct from $q$. Similarly assume that the consecutiveness of $S'$ in $C'$ is certified by the set $S'_e$ of
%Marc-v6
endpoints,
and without loss of generality let $q'_1$ be the endpoint of $q'$ in $S'_e$. Let $r'_1$ denote the other bookend of $S'_e$ with chord $r'$ distinct from $q'$. Observe that either $S_e$ is a factor of 
% MT 08/28/12 - Reviewer noted the over specification of "but not equal to" excludes the case boundary case where S = V or S' = V'.  I don't think we use the fact that they aren't equal, so I'm removing them here and in the next sentence below.
%EG 09/23/12 cancelled Marc's
(but not equal to) 
$q_1C(q_1,q_2)$ or of $C(q_2,q_1)q_1$. 
%Marc-v6: I didn't think you could use "without loss of generality" like this.  I have only ever encountered it when the labels don't mean anything and can be assigned arbitrarily.  Here that is not the case.  
%Eme-v7 you are right, these w.l.o.g. are abusive since sometimes we need reversion, sometimes not. I change these sentences
%Without loss of generality assume the former. 
Assume the former. 
Observe also that either $S'_e$ is a factor of 
(but not equal to) 
$q'_1 C'(q'_1,q'_2)$ or of $C'(q'_2,q'_1) q'_1$. 
%Marc-v6: Same issue as above.
%Without loss of generality assume the former. 
Assume the former. 
%Marc-v6: Do we mean "similar", to be precise? 
%Eme-v7 yes similiar would be much better (symmetric had to be understood as 'up to reversion' and up to changing the operation the 'symmetric one'... not good!).I change the sentence below
%The other cases are symmetric.
We complete the proof under these two assumptions, the other cases are similar.
%

%\xtofComment{I am lost here. We should consider reviewer 1's comment on the case $S=V$ and $S'=V'$. Trying to do so, I do not understand below why $b\notin S_e$. Especially if we have $S=V$ (so $G$ is a permutation graph) and $q$ is the chord of a universal vertex, then it sems that $b\in S_e$ (this is the other bookend of $S_e$).  In fact I don't why we argue that $b$ is not an endpoint of $r$. For me it can be, and there is no problem.}

Let $a$ and $b$ be the first and last endpoints 
%\xtofComment{bookends ?} 
of $C(q_1,q_2)$. Observe that $a\in S_e$ and $b\notin S_e$ and thus $b$ is not an endpoint of $r$. Let $a'$ and $b'$ be the first and last endpoints %\xtofComment{bookends ?} 
of $C'(q'_1,q'_2)$. Observe that $a'\in S'_e$ and $b'\notin S'_e$ and thus $b'$ is not an endpoint of $r'$. By construction, $a$ and $a'$ appear consecutively on $(C,q)~\odot~ (C'^r,q')$. Then
$(S\setminus \{q\}) \cup (S'\setminus \{q'\})$ is consecutive and has bookends $r_1$ and $r'_1$. 
%
%While in $(C,q)\odot (C',q')$, $b$ and $a'$ are consecutive; in $(C,q)~\hat\odot~ (C',q')$, $a$ and $b'$ are consecutive;  and in $(C,q)~\hat\odot~ (C'^r,q')$, $b$ and $b'$ are consecutive. In each of these three cases, if $(S\setminus \{q\}) \cup (S'\setminus \{q'\})$ is consecutive then $a$ or $a'$ is a bookend, 
%It follows that $(C,q)~\odot~ (C'^r,q')$ is the only chord diagram among the four with the required property.
\end{proof}

%\begin{remark}
%\label{rk:unique-circle-join}
%It is easy to refine  Lemma \ref{labelToGraph} and its proof to get that:
%if $C(q_1,q_2)$ and $C(q_2,q_1)$, resp. $C'(q'_1,q'_2)$ and $C'(q'_2,q'_1)$, contain an element which is not an endpoint of an element of $S$, resp. $S'$, then exactly one of the chord diagrams given in the statement of Lemma \ref{labelToGraph} has the required property.
%%%% other formulation :
% With notations of Lemma \ref{labelToGraph}, assume $S_e$, resp. $S'e$, certifies the consecutiveness of $S$, resp. $S'$. Let $q_e$, resp. $q'_e$, be the endpoint of $q$ which is not a bookend of $S_e$, resp $S'_e$. And let $s$, resp. $s'$, be the bookend of $S_e$, resp. $S'_e$, which is not an endpoint of $q$, resp. $q'$. If $q_e$ is not a neigbor of $s$ and $q'_e$ is not a neigbor of $s'$, then it is easy to check that exactly one of the chord diagrams given in the statement of Lemma \ref{labelToGraph} has the required property.
%\end{remark}

\begin{figure}[h]
\begin{center}
%\scalebox{0.5}{\input{reflectionExample.pstex_t}}
\includegraphics[scale=1]{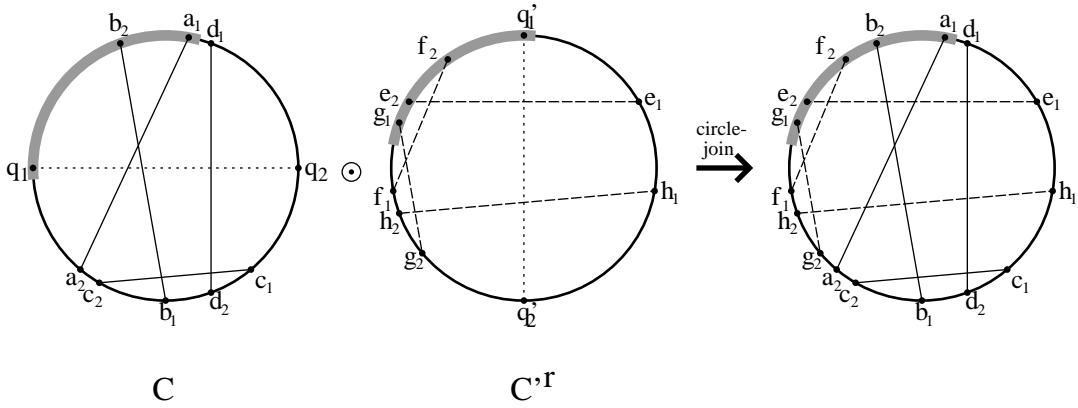}
\end{center}
%\vspace{-0.2in}
\vspace{-0.5cm}
\caption{A consecutivity preserving circle-join requiring a reflection (Lemma \ref{labelToGraph} with $S=\{q,a,b\}$ and $S'=\{q',f,e,g\}$). Here $C'^r$ is the reflection of $C'$ from Figure~\ref{fig:cJoinExample}. The resulting chord diagram is $(C,q)\odot (C'^r,q')$, where $\{a,b,f,e,g\}$ is consecutive with inherited bookends $a$ and $g$.}
\label{fig:reflectionExample}
\end{figure}

%Marc-v6: I don't think this belongs here.  I also thought it asked a bit too much of the reader.  I have moved it to the end of section 2 and reworked it as it appears there.
%As a direct consequence of the fact that the join operation preserves being a circle graph by Lemma \ref{c-join}, and the fact that each grah label is isomorphic to a subgraph of the accessibility graph (Remark \ref{subgraph}), we retrieve the well-known fact that a graph is circle if and only every node of its split-tree is labelled by a circle graph~\cite{easy}. In terms of graph-labelled tree, we obtain:
%\begin{corollary} \cite{easy} 
%\label{cor:circle-labels}
%Let $(T,\mathcal{F})$ is a GLT. The accessiblity graph $\G(T,\mathcal{F})$ is a circle graph if and only if every label graph of $\mathcal{F}$ is a circle graph.
%\end{corollary} 

%----------------------------------------------------------------------------------------------------------------------
%\subsection{Split-tree and chord diagrams}
\subsection{Split-Tree Property}
\label{sec:circle+split}

This subsection shows how a consecutive set of chords/vertices in a chord diagram $C$ of a circle graph $G$ induces a consecutive set of chords/vertices of the chord diagram of the circle graph $G(u)$
%Marc-v6
for
 any node $u$ of the split-tree $ST(G)$. The proof relies on the following 
 %Marc-v6
 result, which can be found in an equivalent form as proposition 9 in~\cite{Cou08}.
 %

%Emeric enonce
%\begin{lemma} \label{circlePartition}
%The split $(A,B)$ induced by any edge of the split-tree of a circle  graph is represented on the chord diagram by four disjoint intervals of endpoints :one pair of opposite intervals for A and one for B
%\end{lemma}  

%Marc-v6: I found the inclusion of "proposition 9" ugly and not easily read.  I think it suffices to cite the paper only.  I made a similar change above.  If you feel strongly, then change it back.  Also, well done finding this result.  I should have read Bruno's paper.
%Eme-v7 as you wish, but this is standard typesetting, at least we should maintain it in the statement below, it is better being part of the statement... Also Bruno is maybe not the first to have proved this result, but he hasn't found it in another place neither
\begin{proposition} [Proposition 9 in \cite{Cou08}] \label{circlePartition}
%\begin{proposition} \cite[Proposition 9]{Cou08} \label{circlePartition}
%\begin{proposition} \cite{Cou08} \label{circlePartition}
%
Let $C$ be a chord diagram for the circle graph $G$.  Let $q$ and $r$ be the extremities of a
% MT 08/28/12 - Reviewer noted that it holds for all tree edges, not just internal tree edges. Why not make it as general as possible?
%internal 
tree-edge in $ST(G)$.  
Then $C$ can be partitioned into four factors $C\sim A_1B_1A_2B_2$ such that 
%Marc-v6: I don't think the notation of the union of two factors has been introduced.  I leave it to you to decide if it needs to explicitly mentioned somewhere, or whether it is "natural enough".
%Eme-v7 it has been defined (implicitely) as a notation for the concatenation of two words (line 5 section 2.4). But I add an emphasized def for this in section 2.4, even if it is now  a little redundant...
$A_1\cup A_2=\bigcup_{x\in L(q)}\{x_1,x_2\}$ and $B_1\cup B_2=\bigcup_{y\in L(r)}\{y_1,y_2\}$. 
\end{proposition}  

Together with Remark~\ref{rem:accessible-split}, the above proposition yields the following:

\begin{corollary} \label{cor:endpoint-A}
Let $q$ and $r$ be the extremities of 
% MT 08/28/12 - Same as above: why not make it as general as possible?
%an internal 
a tree-edge in $ST(G)$.  Let $C\sim A_1B_1A_2B_2$  be a chord diagram for the circle graph $G$ such that $A_1\cup A_2=\bigcup_{x\in L(q)}\{x_1,x_2\}$ and $B_1\cup B_2=\bigcup_{y\in L(r)}\{y_1,y_2\}$. 
% MT 08/31/12 - Reviewer didn't like using symbols in place of English in this case (and one other) and suggested this needed to be checked throughout the paper.  Here the offending use was l \in A(q) instead of l "belongs to" A(q).  I'm not sure I totally agree.  In this case l \in A(q) is a proposition within the logical "iff" statement.  After reviewing this example and the other cited instance, I think the problem arises from mixing English -- "Then the leaf..." -- with the symbols -- l \in A(q).  I have fixed this one.  The other cited instance was commented out due to unrelated changes requested by the first reviewer (I still fixed it, though).  I could not find any other of the "belongs to" type by a ctrl+f search.
Consider an arbitrary leaf $l$ in $ST(G)$.  Then 
%Then the leaf 
$l$ is in $A(q)$ if and only if it has one endpoint in $A_1$ and the other in $A_2$.
\end{corollary}
\begin{proof}
By Remark~\ref{rem:accessible-split}, $A(q)$ and $A(r)$ are the frontiers of the split $(L(q),L(r))$ in $G$. In other words, every leaf of $A(r)$ is adjacent in $G$ to every leaf of $A(q)$. Equivalently, these pairs of leaves correspond to 
%Marc-v6
intersecting
pairs of chords $l, l'$ such that $l\in A_1\cup A_2$ and $l'\in B_1\cup B_2$. Observe that this holds if and only if $l$ (respectively $l'$) has one endpoint in $A_1$ (respectively $B_1$) and the other in $A_2$ (respectively $B_2$).
\end{proof}

If $u$ is a node of $ST(G)$, then applying proposition~\ref{circlePartition} on every tree-edge incident to $u$, we obtain:

\begin{corollary} \label{labelPartitions}
%DGC "enconding" fixed
Let $C$ be a chord diagram encoding the circle graph $G$. If $u$ is a node of degree $k$ in $ST(G)$, then $C$'s endpoints can be partitioned into $2k$ factors $C\sim A_1A_2\dots A_{2k-1}A_{2k}$ such that for every $q$ in $V(u)$, there exists a distinct pair $A_i,A_j$  such that $$A_i\cup A_j=\bigcup_{x\in L(q)}\{x_1,x_2\}.$$
%where $x \in L(q)$ if and only if $x_1,x_2 \in A_i \cup A_j$.
%Moreover, for $q \in V(u)$, every element of $L(q)$ accessible from $q$ has its endpoints in two distinct of these sets.
\end{corollary}

From Corollary \ref{labelPartitions}, given a circle graph with chord diagram $C$ and a node $u$ in its split-tree, we can define 
\emph{the simple chord diagram $\bar C[u]$ of $u$ induced by $C$} as follows: 
%Marc-v6: I thought the following had to be clarified.
%replace in $C$ the factor $A_i$ and $A_j$ containing the endpoints of $L(q)$ by $q$.
%DGC below, replaced "by" with "with"
for each $q \in V(u)$, remove the factors $A_i$ and $A_j$ corresponding to $L(q)$ and replace them with $q$.
%Then $\bar C[u]$ is a simple chord diagram for $u$, which is canonically isomorphic to $\bar C[L]$ for any accessible sample $L$ of $u$ (implying that all these $\bar C[L]$ are isomorphic).
%%, which is not the case in general GLT, see last paragraph below).

%Eme-v7 def below had been taken out of next corollary, but I put it back in the statement, see comment inside
%\bigskip
%For a chord diagram $C$ encoding the circle graph $G$, and a node $u$  in $ST(G)$, we denote
%$$S[u]=\{\  q \in V(u) \ |\  L(q) \cap S \ne \emptyset\ \}.$$

\begin{corollary}\label{graphToLabel}
Let $C$ be a chord diagram encoding the circle graph $G$, and let $u$ be a node in $ST(G)$. 
%Assume 
%Marc-v6
%that
%
%$S$ is a consecutive set of chords in $C$ and let 
%Marc-v6: I don't like introducing notation in a result statement unless it's only going to be used in that result.  Here, this notation seems to be used later in the proof of theorem 3.11.  I think it should therefore have its own definition to help the reader make note of the fact that it will be used in the future, and so that when it is used in the future, they will be able to refer back to more easily.  Who would think to check the statement of a corollary to search for the meaning of some notation they encounter?
%Eme-v7 at first I agreed, and I extracted the def, but now I realize that we give to ddefinitions of the same thing : S[u] here and MP(u) later. So let us define S[u] just locally inside this statement.
%$$S[u]=\{\  q \in V(u) \ |\  L(q) \cap S \ne \emptyset\ \}.$$
%
Assume that $S$ is a consecutive set of chords in $C$.
Let
$$S[u]=\{\  q \in V(u) \ |\  L(q) \cap S \ne \emptyset\ \}.$$
Then $S[u]$ is consecutive in $\bar C[u]$. Moreover if $x$ in $L(q)$ is a bookend for $S$, then $q$ is a bookend for $S[u]$.
\end{corollary}
\begin{proof}
Assume that $C\sim A_1A_2\dots A_{2k-1}A_{2k}$%
%Marc-v6
,
 as defined in Corollary~\ref{labelPartitions}. Without loss of generality,
 %Marc-v6 
 %
 assume that the consecutiveness of $S$ is certified by a factor contained in $A_i\dots A_j$, with $1\leqslant i<j\leqslant 2k$. As $S$ is consecutive, observe that every $A_h$%
%Marc-v6
, $i\leqslant h\leqslant j$,
 corresponds to a distinct marker vertex $q_h$ of $u$. This clearly implies that $S[u]$ is consecutive in $\bar C[u]$. Moreover, as the bookends of $S$ belong to $A_i$ and $A_j$, then the bookends of $S[u]$ are the corresponding marker vertices $q_i$ and $q_j$.
\end{proof}

One can observe that
%Marc-v6
%obviously 
by definition of $\bar C[u]$, 
%Marc-v6
%the conclusion of 
Corollary \ref{graphToLabel} 
%Marc-v6
implies that if $S$ is consecutive in $C$, then $S\cap L$ is consecutive in $C[L]$,
where $L$ is any set of vertices/chords obtained by selecting one accessible leaf in $A(q)$ for every marker vertex $q$ of $u$.

%----------------------------------------------------------------------------------------------------------------------
%\subsection{LBFS and consecutiveness}
\subsection{LBFS Property}
\label{sec:good}

The next theorem is a new structural property of circle graphs. It will be used in Section \ref{sec:charact} to characterize vertex insertion leading to prime circle graphs. 
%The proof supplied below proceeds in stages.  It relies on the concept of module. A \emph{module} is a set of vertices $M$ such that every vertex not in $M$ is either universal to $M$ or isolated from $M$.  A module is \emph{trivial} if $M = V(G)$ or $|M| = 1$. 

\begin{theorem} \label{th:good}
Let $G$ be a prime circle graph. If $x\in V(G)$ is a good vertex of $G$, then $G$ has a chord diagram in which $N(x)$ is consecutive.
\end{theorem}

\begin{proof}
Let $C$ be a chord diagram of $G$. As $G$ is prime, we know that $C$ is unique up to reflection~\cite{Bou87,Cou08}. Assume for contradiction that $N(x)$ is not consecutive in $C$. Let $\sigma$ be an LBFS ordering of $G$ in which $x$ is the last vertex. Let $z$ denote the first vertex in $\sigma$. Either one endpoint of $z$ appears in $C(x_1,x_2)$ and the other in $C(x_2,x_1)$, or the two endpoints 
%Marc-v6
appear
in one of $C(x_1,x_2)$ and $C(x_2,x_1)$. Without loss of generality, suppose that $C(x_1,x_2)$ contains at most one of $z$'s endpoints.

Since $N(x)$ is not consecutive in $C$, at least one vertex/chord has its two endpoints in $C(x_1,x_2)$.  Amongst all such vertices, let $y$ be the one occurring earliest in $\sigma$. Observe that by 
%Marc-v6
construction, 
$y\neq z$. Let $B_y$ be the set of vertices occurring before $y$ in $\sigma$; and let $A_y$ be the set of vertices with the same label as $y$ (including $y$) at the step $y$ is numbered by Algorithm~\ref{alg:LBFS}.

By the choice of $y$, every neighbour $v$ of $y$ such that $v<_{\sigma} y$ has only one of its endpoints in $C(x_1,x_2)$.  Therefore $N(y) \cap B_y \subseteq N(x) \cap B_y$.  It follows that at the step $y$ is numbered we have $label(x)=label(y)$, implying $x \in A_y$.  As $x$ is good, we have $(B_y,A_y)$ is a bipartition of $V(G)$. By 
%Marc-v6
construction, 
$A_y$ contains at least two vertices 
 %Marc-v6
(i.e. 
$x$ and $y$). So if $|B_y|\geqslant 2$, the bipartition $(B_y,A_y)$ defines a split of 
%Marc-v6
$G$,
%
%DGC added the note phrase below.
 contradicting that $G$ is prime. It follows that $B_y=\{z\}$ and $z$ is a universal vertex in $G$; note that $y$ is the second vertex in $\sigma$.

% MT 08/28/12 - Reviewer didn't like the repetition in the argument below.  Rewriting it to borrow more from the preceding paragraph.
%DGC  I've made a few minor adjustments to this paragraph (too many "similar .. above" - and corrected $x \in B_{y'}$ to $x \in A_{y'}$
The same argument as above can be applied to $C(x_2,x_1)$.  There must be a vertex $y'$ both of whose endpoints reside in $C(x_2,x_1)$, and without loss of generality, we can assume $y'$ is the earliest such vertex appearing in $\sigma$.  Observe that $y' \ne z$ and $y' \ne y$, by construction.  Define $B_{y'}$ and $A_{y'}$ similar to $B_y$ and $A_y$ above.  Then following the same argument as above, $N(y') \cap B_{y'} \subseteq N(x) \cap B_{y'}$. Thus $x \in A_{y'}$.  But now both parts of the bipartition $(B_{y'},A_{y'})$ have size at least two (recall that $y,z \in B_{y'}$).  It follows that $(B_{y'},A_{y'})$ is a split, contradicting $G$ being prime.
%Marc-v6
%Again,
%
% as $N(x)$ is not consecutive in $C$,  at least one vertex/chord has its two endpoints in $C(x_2,x_1)$. Amongst all such vertices, let $y'$ be the one occurring earliest in $\sigma$. Observe that by 
 %Marc-v6
% construction,
 %
% $y'\neq z$ and $y'\neq y$. 
%Marc-v6
%As for vertex $y$, we 
%We define $B_{y'}$ to be the set of vertices occurring before $y'$ in $\sigma$.  Notice that $B_{y'}$ must include $z$ and $y$, by construction.  We denote $A_{y'}$ to be the set of vertices
%
% with the same label as $y'$ (including $y'$) at the step $y'$ it is numbered by Algorithm~\ref{alg:LBFS}. By the choice of $y'$, every neighbour $v'$ of $y'$ such that $v'<_{\sigma} y'$ has only one of its endpoints in $C(x_2,x_1)$.  Therefore $N(y') \cap B_{y'} \subseteq N(x) \cap B_{y'}$.  It follows that at the step $y'$ is numbered we have $label(x)=label(y')$, implying $x \in B_{y'}$. But now both parts of the bipartition $(B_{y'},A_{y'})$ has size at least $2$
 %Marc-v6
% (recall $y,z \in B_{y'}$). It follows that $(B_{y'},A_{y'})$ is a split, contradicting $G$ being prime.
 %
\end{proof}

%----------------------------------------------------------------------------------------------------------------------
\subsection{Good Vertex Insertion in Circle Graphs}
%\subsection{LBFS incrmental characterization of prime circle graphs}
\label{sec:charact}

We now present 
%Marc-v6
an
LBFS incremental characterization of prime circle graphs. That is, assume that adding a new vertex $x$ to a circle graph $G$ yields a prime graph $G+x$.
%Marc-v6: Would need a separate sentence here, not a comma.  But I suggest rewording as follows afterward.
%, which conditions are required on $ST(G)$ to prove that $G+x$ is a circle graph as well? 
We answer the following question: which properties of $ST(G)$ are required for $G+x$ to be a circle graph as well?  
%Eme-v7
%We answer the question 
We use
the results from the three previous subsections and the incremental charaterization of the split decomposition in \cite{GPTC11a}.

We first need some definitions from \cite{GPTC11a}. Let $G$ be an abitrary (connected) graph
%Marc-v6
%$G$ and a
and consider some subset
$S\subseteq V(G)$. 
% MT 08/28/12 - Reviewer wanted stamps to apply not just to split-tree but GLTs in general because later we use the labels on the intermediate GLTs formed during the insertion of a new vertex.
 Let $(T,\mathcal{F})$ be a GLT such that $Gr(T,\mathcal{F}) = G$.  We stamp the marker vertices of $(T,\mathcal{F})$ with respect to $S$ as follows. 
 %
% We stamp the marker vertices of the split-tree $ST(G)$ with respect to $S$ as follows. 
 If $q$ is a marker vertex opposite a leaf $l\in S$, 
%Marc-v6: same as before with "respectively" in brackets as I have always seen it.
(respectively $l\not\in S$) 
%in $ST(G)$, 
we say that $q$  is \emph{perfect} (respectively \emph{empty}). 
Let $q$ be a marker vertex not opposite a leaf. Then $q$ is 
\emph{perfect} if $S\cap L(q)= A(q)$; \emph{empty} if $S\cap L(q)= \emptyset$; and \emph{mixed} otherwise.
Let $P(u)$ denote the set of perfect marker vertices of the node $u$, 
and let $MP(u)$ denote the set of mixed or perfect (i.e. non-empty) marker vertices of the node $u$ in $ST(G)$: i.e. $MP(u)=\{q\in V(u)\mid S\cap L(q)\not= \emptyset\}$. 
%Note than an edge is a $PE$ edge if one extremity is perfect and the other is empty; similar terminology is used for other states.
%Eme-v7: def above removed, confusing with the notation MP used for a SET, not an edge, and PP PE used only in a forthcoing theorem from the split paper.

\begin{lemma} \label{mixed}
Let $G=(V,E)$ be a circle graph and 
%Marc-v6
let 
$C$ be a chord diagram  of $G$ in which the set $S\subseteq V$ is consecutive. If $q$ is a mixed marker vertex of the node $u$ in $ST(G)$ marked with respect to $S$, then $L(q)$ contains a leaf $\ell$ that is a bookend of $S$.
\end{lemma}

%\com{*** the proof below has been rewritten, I did not understand marc's one, which I guess was possibly false, since it seemed to prove that $A_i\subseteq S_e$ was impossible as soon as $A_i\cap S_e\not=\emptyset$, whereas I guess that $A_i$ could contain one only endpoint  ***}

\begin{proof}
By Corollary~\ref{labelPartitions}, there is a pair of factors $A_i$ and $A_j$ in $C$ such that $y \in L(q)$ if and only if 
%Marc-v6: Same subset notation issue as before.
%Eme-v7 I don't understand this comment
$y_1,y_2 \in A_i \cup A_j$. 
Let $S_e$ be a set of endpoints certifying that $S$ is consecutive in $C$. Since $q\in MP(u)$, we know $L(q) \cap S \ne \emptyset$.  Therefore $S_e \cap A_i \ne \emptyset$ or $S_e \cap A_j \ne \emptyset$ (or both). Assume without loss of generality that $S_e \cap A_i \ne \emptyset$.  

If $A_i$ does not contain a bookend of $S_e$, this implies that $A_i\subset S_e$. Therefore every chord with one endpoint in $A_i$ has its other endpoint in
%Marc-v6
$A_j$, by definition of $S_e$ being consecutive.
By Corollary~\ref{cor:endpoint-A}, $A(q)$ is the set of chords with 
%Marc-v6
exactly 
one endpoint in $A_i$. It follows that $A_j$ cannot be a subset of $S_e$: 
%Marc-v6: Simplifying/clarifying argument/language for reader; please check that I'm correct.  If you don't agree that it is simpler, then change back.
%if $y_1$ is an endpoints in $A_i$, as it belongs to $S_e$, the other endpoint $y_2$ of the chord $y$ lies in $A_j$ and does not belong to $S_e$.  Moreover as $q$ is mixed, $A(q)\neq S\cap L(q)$. As $A_i\subset S_e$, $A(q)\subset S$ and so there exists $z\in (S\cap L(q))\setminus A(q)$. Observe then that the two endpoints $z_1$ and $z_2$ of $z$ belong to $A_j$. But only one of them belongs to $S_e$. It follows that $A_j$ contains a bookend of $S_e$.
%Eme-v7 I leave this job to the next reader, maybe Christophe who had rewritten this proof
%%%TO-BE-DONE : check the new end of the proof
if it were, then $A_i \cup A_j \subseteq S_e$, and so there would be a chord with both its endpoints in $S_e$, a contradiction by definition of $S_e$ being consecutive.  We also can not have $A_j \cap S_e = \emptyset$: 
%xtof-v8: minor correction
%if we did, then we would have $A_i \cap S_e = A_i$, meaning $L(q) \cap S = A(q)$, implying that $q$ is perfect, a contradiction.  Thus, $A_j \cap S_e \ne \emptyset$ but $A_j \not \subseteq S_e$.  It follows that $A_j$ contains a bookend of $S_e$.
if so, we would have $(A_i\cup A_j) \cap S_e = A_i$, then by Corollary~\ref{cor:endpoint-A} and the consecutiveness of $S$, $L(q) \cap S = A(q)$ which implies that $q$ is perfect, a contradiction.  Thus, $A_j \cap S_e \ne \emptyset$ but $A_j \not \subseteq S_e$.  It follows that $A_j$ contains a bookend of $S_e$.

\end{proof}

We extract the following result for arbitrary graphs from \cite{GPTC11a}:

\begin{theorem} [Theorem 4.21 in \cite{GPTC11a}]
\label{prop:ST-prime}
A graph $G+x$ is a prime graph if and only if $ST(G)$ marked with respect to $N(x)$ satisfies the following:
\begin{enumerate}
\item Every marker vertex not opposite a leaf is mixed,
\item Let $w$ be a degenerate node. If $w$ is a star node, the centre of which is perfect, then $w$ has no empty marker vertex and at most 
%Marc-v6: to clarify, please make sure this is the correct interpretation.  The fact that I'm not sure suggests it needs clarification.
%Eme-v7 ok
one other perfect marker vertex; 
%
% MT 08/28/12 - Reviewer correctly noted ambiguity of "otherwise".  To what does it refer: the star or the star whose centre is perfect?
and in all other cases,
%otherwise 
$w$ has at most one empty marker vertex and 
%Marc-v6: do we need this as well?
%Eme-v7: not much, but let us state completely the characterization for cultural purpose !
at most
one perfect marker vertex.
\end{enumerate}
\end{theorem}

\begin{figure}[h] %[htbh]
\begin{center}
\includegraphics[scale=1]{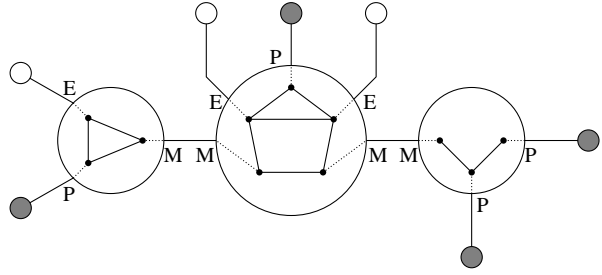}
\end{center}
\vspace{-0.2in}
\caption{States -- P for ``perfect'', M for ``mixed'' and E for ``empty'' -- assigned to marker vertices with respect to the shaded leaves representing $N(x)$. This split-tree satisfies Theorem \ref{prop:ST-prime}.}
\label{fig:stateExample}
\end{figure}

%One can observe that $MP(u)$ is thus in bijection with any accessible sample of $u$.
See Figure \ref{fig:stateExample} for an example satisfying the conditions of Theorem \ref{prop:ST-prime}.

%%%TO-BE-DONE : check the theorem's proof below (the main one of the paper... needs to be perfect)

\begin{theorem} \label{splitTreeTheorem}
Let $G+x$ be a prime graph such that $x$ is a good vertex and $G$ is a circle graph.
Then $G+x$ is a circle graph if and only if for every node $u$ in $ST(G)$, marked with respect to $N(x)$, $G(u)$ has a chord diagram in which $MP(u)$ is consecutive, with the mixed marker vertices being bookends.
\end{theorem}

\begin{proof} 
\emph{Necessity:}  If $G+x$ is a  circle graph, it has a  chord diagram $C'$. By Theorem~\ref{th:good}, $N(x)$ is consecutive in $C'$. 
%Marc-v6
%Thereby
Therefore
 $N(x)$ is consecutive in the chord diagram $C=C'[V(G)]$ of $G$. 
%Observe that, for every node $u$ of $ST(G)$, 
 %Marc-v6: moving it later for clarity: don't wait too long to introduce the main part of the sentence.  Better to have the subordinate clauses afterwards.
%%%%%%%%%% %by definition of $MP(u)$, 
% we have 
%Marc-v6: I found this notation difficult to understand and hard to locate where it was introduced, which seems to be corollary 3.7, and you can refer to my comment there about that.  Can we find some other way to express the same concept, or at least give a separate definition, and then refer to that definition (e.g. in parentheses) when using it below?
%
%$MP(u)=N(x)[u]$, 
%
%by definition of $MP(u)$.
%Eme-v7 I share your concern above, so I propose to detail more this equality in this proof. See below.
%xfot-v8: minor corrections
Let $u$ be a node of $ST(G)$, which by assumption is marked  with respect to $N(x)$.
%By definition of $MP(u)$ (preceding Lemma~\ref{mixed}), we have $MP(u)=\{q\in V(u)\mid N(x)\cap L(q)\not= \emptyset\}$.
%By definition of $N(x)[u]$ (inside Corollary \ref{graphToLabel}), we have $N(x)[u]=\{\  q \in V(u) \ |\  L(q) \cap N(x) \ne \emptyset\ \}.$
%That is, for every node $u$ of $ST(G)$, $MP(u)=N(x)[u]$.
By the definition of $MP(u)$ (preceding Lemma~\ref{mixed}) and of $N(x)[u]$ (inside Corollary \ref{graphToLabel}), 
we have $MP(u)=N(x)[u]=\{\  q \in V(u) \ |\  L(q) \cap N(x) \ne \emptyset\ \}.$
So, according to Corollary~\ref{graphToLabel}, $MP(u)$ is consecutive in $C[u]$, a chord diagram of $G(u)$.  By Lemma~\ref{mixed}, if $q$ is a mixed marker vertex of $u$, then $L(q)$ contains a leaf that is a bookend of $S$, which 
%Marc-v6
implies, by Corollary~\ref{graphToLabel},
 that $q$ is a bookend of $MP(u)$.

%Eme-v7  proof below rewritten with better organization of the logical implications and with a more rigorous induction. Sorry I had to delete the older proof, it had become too unreadable.

\emph{Sufficiency:} 
 By assumption, $ST(G)$ satisfies the following property:
 (A) \emph{for every node $u$, $G(u)$ has a chord diagram $C_u$ in which $MP(u)$ is consecutive with mixed marker vertices being bookends}. 
%Marc-v6
By Theorem~\ref{prop:ST-prime}, the extremities of every internal tree-edge of the split-tree $ST(G)$
%Marc-v6
are
 mixed.
Hence $ST(G)$ also satisfies the following property:
(B) \emph{for every internal tree-edge $e=uu'$, the extremities $q\in V(u)$ and $q'\in V(u')$
of $e$  are bookends of $MP(u)$ and $MP(u')$ respectively}.

Also, one can observe that the internal tree-edges of $ST(G)$ form a path. Indeed, because 
a consecutive set of chords has two distinct bookends, each node $u$ has at most two mixed marker vertices, and hence it has at most two neighbours in $ST(G)$.

%By Theorem \ref{prop:ST-prime} and by definition of perfect marker vertices, 
By definition of perfect marker vertices, $ST(G)$ also satisfies the following property: (C) \emph{$N(x)$ is the set of leaves whose opposite marker vertices belong to $P(u)$ for some node $u$}.
For a node $v$ in a GLT obtained by a series of node-joins from $ST(G)$, let us extend the previous definitions and denote $P(v)$ to be the set of marker vertices of $v$ opposite a leaf belonging to $N(x)$, and $MP(v)$ this set together with mixed marker vertices, defined as extremities of internal edges.

We now prove that $G$ has a chord diagram $C$ in which $N(x)$ is consecutive, by induction on the number of nodes of a GLT of $G$ satisfying satisfying properties (A), (B) and (C). This 
%Marc-v6
would obviously imply
that $G+x$ is a circle graph. If %$ST(G)$
such a GLT has a unique node $u$, then $N(x)$ is the set of leaves opposite marker vertices in $MP(u)=P(u)$, and the result trivially holds with $C$ isomorphic to $C_u$. 
Assume that the result holds for every such GLT with $k$ 
nodes, and consider a GLT with $k+1$ 
nodes satisfying properties (A), (B) and (C). 
Let $e=uu'$ be an internal tree-edge with extremities $q$ and $q'$ and let $C_{u}$, $C_{u'}$ be two respective chord diagrams witnessing the consecutiveness of $MP(u)$ and $MP(u')$. 
By Lemma~\ref{labelToGraph}, 
%Marc-v6: Changing phrase ordering.
the set $(MP(u)\setminus\{q\})\cup(MP(u')\setminus\{q'\})$ is consecutive, inheriting its bookends from $MP(u)$ and $MP(u')$, in (at least) one of the following chord diagrams:
$$(C_u,q)\odot (C_{u'},q'),~~~~~(C_u,q)~\hat\odot~ (C_{u'},q'),~~~~~(C_u,q)\odot (C_{u'}^r,q'),~~~~~(C_u,q)~\hat\odot~ (C_{u'}^r,q').$$ 

That chord diagram encodes the graph $G(v)$ resulting from the join between $G(u)$ and $G(u')$ with respect to $q$ and $q'$, by Remark~\ref{c-join}. 
This yields a new GLT
 for $G$ (recall the definition of node-join) with $k$  
nodes. 
By definition,
$MP(v)=(MP(u)\setminus\{q\})\cup(MP(u')\setminus\{q'\})$, which
 is consecutive in this chord diagram of $G(v)$, with mixed marker vertices being bookends. Hence properties (A) and (B) are satisfied by this GLT.
 And we have also $P(v)=P(u)\cup P(u')$. Hence property (C) is also satisfied by this GLT.
 
 We already proved that $ST(G)$ satisfies properties (A), (B) and (C), hence the result.
\end{proof}

\begin{figure}[htbh]
\begin{center}
\includegraphics[scale=1.2]{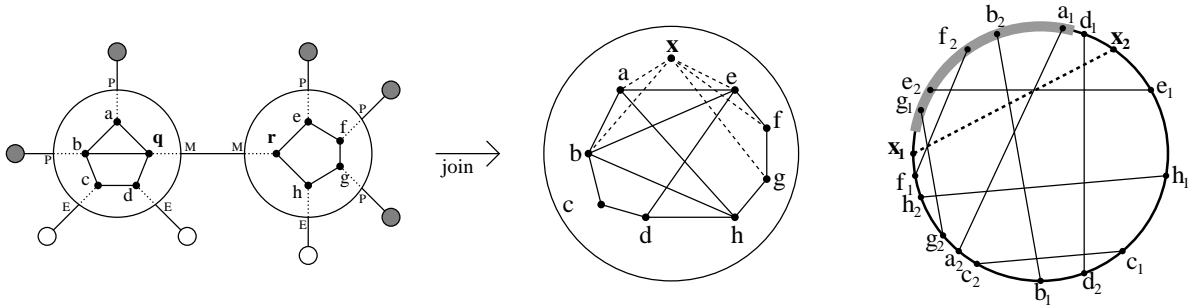}
\end{center}
\vspace{-0.5cm}
\caption{Example of insertion of $x$. The letters P, E, M stand for perfect, empty, mixed, respectively. The involved consecutivity preserving circle-join is the one illustrated in Figure \ref{fig:reflectionExample}.}
\label{fig:states-examples}
\end{figure}

Figure \ref{fig:states-examples} provides an example of 
%Marc-v6
the insertion of a vertex $x$ to a circle graph $G$, where $N(x)=\{a,b,e,f,g\}$ and $G$ is described by the chord diagrams in Figures 
%Marc-v6: Make the numbers consecutive by reordering references
\ref{fig:cJoinExample} and \ref{fig:reflectionExample}.
%

%Marc-v6: why the skip?
%Eme-v7 to separate the concluding paragraph below from the example paragraph above, they are really independent in the reading
\medskip
%

%Marc-v6: The sentence's length is far too heavily weighted toward the first part.  I also think it would help the reader to reinforce what was done in that proof, since it's so important to what follows.  Better to make it explicit rather than rely on them to figure out what are the important points from that construction.  I propose the rewrite below.
%The construction detailed in the proof of the sufficient condition of Theorem \ref{splitTreeTheorem}, consisting in successive circle-join operations preserving consecutiveness and bookends  (Lemma \ref{labelToGraph}) along a path of tree-edges, 
% is the key to the recognition algorithm detailed in next Section \ref{sec:algorithm}.
The construction applied in the proof of sufficiency for Theorem~\ref{splitTreeTheorem} is the basis of our circle graph recognition algorithm.  Recall that successive circle-joins were applied to a path of labels in the split-tree, and each of these circle-joins preserved consecutiveness and bookends.  The next section shows how that construction is used to recognize circle graphs.

\section{Circle Graph Recognition Algorithm} 
\label{sec:algorithm}

We now have the material to present our circle graph recognition algorithm. It relies on the split decomposition algorithm of~\cite{GPTC11a} and inserts the vertices one at a time according to an LBFS ordering $\sigma=x_1<\dots< x_n$. 
% MT 08/28/12 - Reviewer noted that none of this description of the split data-structure is ever used and recommend we get rid of it.  Adding next sentence in its place.
Implementation details from it that are needed for this paper will be introduced as required in the sections that follow.
%For the correctness 
%Marc-v6
%issue 
%of the algorithm, we 
%Marc-v6
%will need to assume that, as each vertex $x_i$ is inserted
%
% in $G_{i-1}=G[\{x_1,\dots,x_{i-1}\}]$, the split decomposition algorithm maintains a data-structure for $ST(G_i)$ which
%Marc-v6
%is basically: a pointer representation of the tree structure of a GLT;
% 
%an adjacency list encoding every prime node label; the 
%Marc-v6
%type (i.e. clique or star) of every degenerate node
%
%(in case of star, the centre is distinguished); 
%and some LBFS information at every node (e.g. the last inserted marker vertex). 
%Marc-v6: I don't think that description adds anything or is needed anywhere else in the paper.  I've deleted it (see my comment there).
%A more precise decription of the parts of this data-structure used in the present paper is given in Section \ref{sub:PCCD}.
%Marc-v6: reorder to make it "smoother".
The reader should refer to~\cite{GPTC11a} for complete implementation details.
%

%Marc-v6
For circle graph recognition, we will additionally need to maintain, at each prime node, a chord diagram.  
This is not required for degenerate 
%Marc-v6
nodes
 as their (potentially many) chord diagrams all have the same generic 
%Marc-v6
structure
 (see Figure~\ref{fig:degenerateExample}). 
 %Marc-v6: Clarify what is meant by the sentence below.
 %Nevertheless appropriate chord diagrams for degenerate nodes are built in time. 
Whatever chord diagrams for degenerate nodes are required by the algorithm will be constructed as needed.
%

%The precise way we handle this will be described in Section~\ref{sec:implementation+DS}.

We first briefly describe how the split decomposition algorithm of~\cite{GPTC11a} updates the split-tree of a graph under a vertex insertion. Based on 
%Marc-v6
this,
 we outline the vertex insertion test for circle 
 %Marc-v6
 graph recognition
  and prove its correctness. The data-structure and complexity issues are postponed to Section~\ref{sec:implementation+DS}.

%----------------------------------------------------------------------------------------------------------------------
\subsection{Incremental Modification of the Split-Tree}
\label{sub:incremental-ST}

%This subsection suming up the general algorithm from \cite{GPTC11a} can be omitted, unless  for completeness of the present paper.

%Marc-v6: Rewritten for clarity.
%This subsection sums up the general algorithm from \cite{GPTC11a} for completeness of the present paper.The specific cases and features of this algorithm that need refinements in the circle graph case are detailed in the next subsection.
This subsection summarizes the general algorithm from~\cite{GPTC11a}.  The next subsection details specific cases and features of that algorithm that will be modified for the purposes of recognizing circle graphs.
%

%Marc-v6
%DGC added (marked ..)
We say that a node $u$ in a GLT (marked with respect to some set of leaves $S$)
is \emph{hybrid} if every marker vertex $q\in V(u)$ is either perfect or 
%Marc-v6
empty,
 and its opposite is mixed. 
 % MT 08/28/12 - Split paper reviewer wanted definition changed to "maximal fully mixed subtree."  Adding the corrected definition from that paperbelow.
 % EG 09/23/12 - I dislike this heavy 'maximal', we do not need to obey, let me remove it and let us discuss
%A \emph{maximal fully-mixed subtree} 
A \emph{fully-mixed subtree} 
$T'$ of a GLT $(T,\mathcal{F})$ is a subtree of $T$ such that: it contains at least one tree-edge;
the two extremities of all its tree-edges are mixed;
%all of its tree edges are fully-mixed, 
and it is maximal for inclusion with respect to these properties. 
% A subtree of a GLT $(T,\mathcal{F})$ is \emph{fully-mixed} if the extremities of its tree edges are all mixed and if it is maximal for 
%Marc-v6
%this
%
%property. 
For a degenerate node $u$, we denote:
\vspace{-0.2cm}
\begin{eqnarray*}
P^*(u) & =  & \{ q \in V(u) \mid q \textrm{ perfect and not the centre of a star} \}, \\
E^*(u) & = & \{ q \in V(u) \mid q \textrm{ empty, or } q \textrm{ perfect and the centre of a star} \}. 
\end{eqnarray*}

\begin{theorem} [Theorem 4.14 in \cite{GPTC11a}]%[Incremental characterization] 
\label{th:cases}
Let $ST(G) = (T,\mathcal{F})$ be marked with respect to a subset $S$ of leaves. Then exactly one of the following conditions holds:

%DGC:  I've introduced $u$ and $e$ and $T$ in the text and made some other minor changes
%DGC2 changed the u, q, T in all 7 cases.
\begin{enumerate}
\item $ST(G)$ contains a clique node $u$ whose marker vertices are all perfect, and this node is unique; %$u$ 
%the incident tree edges to which are the set of tree edges with two perfect extremities;
\item $ST(G)$ contains a star node $u$ 
whose marker vertices are all empty except the centre, which is perfect, and this node is unique;
%the incident tree edges to which are the set of tree edges with one perfect extremity and other empty;
\item $ST(G)$ contains a unique hybrid node $u$ and this node is prime;
\item $ST(G)$ contains a unique hybrid node $u$ and this node is degenerate;
%%%TO-BE-DONE check my english in the two following sentences
%Eme-v7 replaced PP and PE with words (this PP PE was used only here), please check
%\item $ST(G)$ contains a $PP$ tree edge, %$e$ 
\item $ST(G)$ contains a tree-edge $e$ whose extremities are both perfect and this edge is unique;
%\item $ST(G)$ contains a $PE$ tree edge, %$e$ 
\item $ST(G)$ contains a tree-edge $e$ with one extremity perfect and the other empty and this edge is unique;
\item $ST(G)$ contains a unique fully-mixed subtree $T$.
\end{enumerate}
\end{theorem}

%If $ST(G) = (T,\mathcal{F})$ is a GLT marked with respect to a subset $S$ of
%Marc-v6
%its
%
%leaves, then it contains exactly one among the following :
%
%
%\begin{enumerate}
%\item  a unique clique node 
%Marc-v6
%$u$,
%
% the marker vertices of which are all perfect;
%\vspace{-0.2cm}
%\item  a unique star node 
%Marc-v6
%$u$,
% 
%the marker vertices of which are all empty except the centre, which is perfect;
%\vspace{-0.2cm}
%\item  a unique hybrid node $u$, which is prime;
%\vspace{-0.2cm}
%\item  a unique hybrid node $u$, which is degenerate;
%\vspace{-0.2cm}
%\item  a unique tree edge
%Marc-v6
%$e$,
%
%both of whose extremities are perfect;
%\vspace{-0.2cm}
%\item  a unique tree edge 
%Marc-v6
%$e$, one of whose extremities is perfect, the other empty;
%\vspace{-0.2cm}
%\item  a unique fully-mixed subtree.
%\end{enumerate}
%\end{theorem}

%Moreover, in every case, the unique node/edge/subtree is obtained from $T$ by deleting, for every tree edge $e$ with a perfect or empty extremity $q$ whose opposite $r$ is mixed, the tree edge $e$ and the node or leaf corresponding to $r$. %In case 1 and case 2, the node, together with its adjacent edges, is obtained in this way.

%Now letting $S=N(x)$, for $x$ a new vertex, the way $ST(G)$ has to be modified to obtain $ST(G+x)$ can be described. For the first 6 cases it works as follows:

%
Now, for a new vertex $x$, and letting $S=N(x)$, 
the way $ST(G)$ has to be modified to obtain $ST(G+x)$ can be described as follows.

% MT 08/28/12 - Reviewer noted that u isn't defined below.  I disagree that we need to define it as each time it is implicit in the reference to the appropriate case in the theorem.  Besides, I tried introducing u into the statements below and it just ruined their flow.

%\xtofComment{I think we need to state what $u$ is (as well as what $e$ is in case 5 and 6. What about adding\\ " can be described as below (we denote by $u$ / $e$ respectively  to the unique vertex / unique tree edge identified in the different cases of Theroem)" ?}

% EG 09/23/12 I agree with Christophe's proposal, senence added above

\begin{itemize}
\item If one of cases 1, 2 and 3 of Theorem~\ref{th:cases} holds, then $ST(G+x)$ is obtained by  adding to $u$ a marker vertex $q$ adjacent in $G(u)$ to 
%Marc-v6
precisely
$P(u)$ and making the leaf $x$ the opposite of~$q$.
\item If case 4 of  Theorem~\ref{th:cases} holds, then $ST(G+x)$ is obtained in two steps:

\begin{enumerate}
\item performing the node-split corresponding to 
%Marc-v6
$(P^*(u),E^*(u))$,
thus creating a tree-edge
%Marc-v6
$e$,
 the extremities of which are perfect or empty;
\item subdividing $e$ with a new ternary node $v$ adjacent to $x$ and $e$'s extremities, such that $v$ is a clique if both extremities of $e$ are perfect, and 
otherwise $v$ is a star whose centre is 
%Marc-v6
the
opposite of $e$'s empty extremity.
\end{enumerate}

\item If case 5 of  Theorem~\ref{th:cases} holds, then $ST(G+x)$ is obtained by subdividing $e$ with a new clique node adjacent to 
%Marc-v6: change order to remove possible ambiguity that they are BOTH x and e's extremities
$e$'s extremities and $x$.
\item If case 6 of  Theorem~\ref{th:cases} holds, then $ST(G+x)$ is obtained by subdividing $e$ with a new star node adjacent to 
%Marc-v6: same order change as above.
$e$'s extremities and $x$,
 such that the centre of the star is opposite $e$'s empty extremity.

%We can observe that in each of the first 6 cases, except case 3, the modification amounts to update or create a new degenerate node. Thanks to Corollary~\ref{cor:circle-labels}, this has no impact for the circle graph recognition problem. These modifications will be thereby handle by the split decomposition algorithm as described in~\cite{GPTC11a}. Case 3 amounts to update a prime node $u$ by attaching a new leaf to it. In other words, adding a new vertex to the prime circle graph $G(u)$ yields a new prime graph $G'(u)$. We need to check whether $G'(u)$ is circle as well. As the vertex insertion ordering is an LBFS ordering, thanks to Theorem~\ref{splitTreeTheorem} and Lemma~\ref{inducedLBFS}, the necessary and sufficient condition is that the neighbourhood of the new vertex is consecutive in the chord diagram of $G(u)$. 
%%%%This is easy to check with the appropriate data-structure (see Section~\refsec:implementation+DS).

%The core of the circle graph recognition algorithm resides in dealing with the seventh case of Theorem~\ref{th:cases}. Indeed when $ST(G)$ contains a fully-mixed subtree, a new prime node is built in $ST(G+x)$ from existent nodes of $ST(G)$. This will involve  a series of circle-join operations. Let us first recall from~\cite{GPTC11a} how the split-tree is updated in that case. So if case 7 of  Theorem~\ref{th:cases} holds, then $ST(G+x)$ is obtained in three steps:
%
\item If case 7 of  Theorem~\ref{th:cases} holds, then $ST(G+x)$ is obtained in three steps:
\begin{enumerate}
\item {[\emph{cleaning step}]} 
%Marc-v6
performing,
for every degenerate node $u$ of $T$
%DGC removed "the fully-mixed subtree of $ST(G)$," 
the node-splits defined by $(P^*(u),V(u) \setminus P^*(u))$ and/or $(E^*(u),V(u) \setminus E^*(u))$ as soon as they are splits of $G(u)$. The resulting GLT is denoted $c\ell(ST(G))$, for \emph{cleaned} split-tree.
\item {[\emph{contraction step}]} contracting, by a series of node-joins,  the fully-mixed  subtree of $c\ell(ST(G))$ into a single node~$u$;% (marked by $N(x)$)
\item {[\emph{insertion step}]} adding to node $u$ a marker vertex 
%Marc-v6
$q_x$, 
adjacent in $G(u)$ to 
%Marc-v6
precisely
%Marc-v6
$P(u)$, and making $q_x$ opposite $x$. 
The resulting node $u$ is prime. 
\end{enumerate}

\end{itemize}

This combinatorial charaterization of $ST(G+x)$ from $ST(G)$ is valid with no assumption on $x$. 
%Marc-v6: "for the sake of complexity" doesn't quite work, since "complexity" has a different meaning in everyday English.  You would have to say something like "complexity analysis".  I suggest a reworking as below.
%For sake of complexity, it is implemented for new vertices inserted with respect to a LBFS ordering. See \cite{GPTC} for all the details.
The split decomposition algorithm in~\cite{GPTC11a} applies this characterization but inserts vertices with respect to an LBFS ordering because doing so 
%Eme-v7: I changed "facilitates" into "allows" below, since LBFS in crucial for complexity
%facilitates 
allows
its efficient implementation.

\subsection{Incremental Circle Graph Recognition Algorithm}

Here we describe how 
%Marc-v6
to
 refine the general construction described in Subsection \ref{sub:incremental-ST} 
%Marc-v6
for the purposes of recognizing circle graphs.
 Let us repeat again that, while inserting vertices according to an LBFS ordering, we maintain the split-tree of the input graph as in \cite {GPTC11a} and 
%Marc-v6
%moreover 
with each prime node we associate a chord diagram.
%
%The vertex insertion procedure in a circle graph is outlined below as Algorithm~\ref{alg:vertexinsertion}.
%
%%%The vertex insertion procedure is outlined as Algorithm~\ref{alg:vertexinsertion}. Let us repeat again that, while inserting vertices according to an LBFS ordering, we maintain the split-tree of the input graph and with each prime node we associate a chord diagram.
%
%Let $G$ be a circle graph, and let $S=N(x)$, for $x$ a new vertex, the way $ST(G)$ has to be modified to obtain $ST(G+x)$ can be described.
%
Let $G$ be a circle graph, and
%Marc-v6: similar issue as before but not as bad: referring to x in N(x) before it's been introduced
for a new vertex $x$, let $S=N(x)$.  
%Marc-v6: awkward phrasing in English
%We consider how the modification of $ST(G)$ to get $ST(G+x)$ involves chord diagram updates.
%Eme-v7 isn't it 'necessitate' below instead of 'necessitates'?
We consider how the changes to $ST(G)$ in arriving at $ST(G+x)$ %necessitates 
necessitate
updates to the chord diagrams being maintained at prime nodes.

\begin{itemize}

\item  If one of cases 1, 2, 4, 5 or 6 of Theorem~\ref{th:cases} holds, then 
the 
%Marc-v6
changes to $ST(G)$ amount to updating a degenerate node or creating a new degenerate node.  By
Corollary~\ref{cor:circle-labels}, this has no impact 
%Marc-v6
on
the circle graph recognition problem. These modifications will 
%Marc-v6: Your use of "thereby" throughout the document is incorrect.  It is not synonymous with "therefore", "consequently", etc.  It is only used in cases where something arises indirectly by means of another process.  For example, if you prove A, but in the process end up proving B, then you have "proved A, thereby proving B".    
therefore be handled by 
the split decomposition algorithm as described in~\cite{GPTC11a}. 

\item Case 3 of Theorem~\ref{th:cases} amounts to 
%Marc-v6
updating
 a prime node $u_x$ by attaching a new leaf to it. In other words, adding a new vertex to the prime circle graph $G(u_x)$ yields a new prime graph. We need to check whether this new prime graph is a circle graph as well. As the vertex insertion ordering is an LBFS ordering, 
%Eme-v7 isn't a comma missing after LBFS ordering above ? I add it
 %Marc-v6: "thanks to" is too colloquial, plus move the important part of the sentence earlier before introducing the theorems, which seem to relate to nothing at the time they are introduced.
 %thanks to 
 the necessary and sufficient condition is that the neighbourhood of the new vertex is consecutive in the chord diagram of $G(u_x)$ (Theorem~\ref{splitTreeTheorem} and Lemma~\ref{inducedLBFS}).

\item
The core of the circle graph recognition algorithm resides in 
%Marc-v6
%dealing with 
case  7 of Theorem~\ref{th:cases}. 
%Indeed 
When $ST(G)$ contains a fully-mixed subtree, a new prime node is built in $ST(G+x)$ from 
%Marc-v6
existing nodes of $ST(G)$.
%
%This will involve  a series of circle-join operations. Let us first recall from~\cite{GPTC} how the split-tree is updated in that case. So if case 7 of  Theorem~\ref{th:cases} holds, then $ST(G+x)$ is obtained in three steps:
%

%\begin{enumerate}

%\item
The first step
%Marc-v6
in case 7, 
namely the cleaning step, works as in the split-tree algorithm of~\cite{GPTC11a}. 
It produces the GLT 
%Marc-v6
$cl(ST(G))$, the fully-mixed tree that will be transformed, in the second step, 
into a single node by means of all possible node-join operations.
If $u_x$ is the result of the above node-joins, then let $G(u_x) + x$ be the prime graph obtained in the third step by adding a vertex (corresponding to $x$) to $G(u_x)$ with neighbourhood $P(u_x)$.  

Notice the similarity between the node-joins in the second step and the construction in the proof of sufficiency of Theorem~\ref{splitTreeTheorem}.  In order to apply that theorem, we make the following observation:

\hfil
\vbox{
\hsize=14cm
\parskip=0mm
\it
the fully-mixed subtree of $cl(ST(G))$, as considered in \cite{GPTC11a},
corresponds canonically to $ST(G(u_x))$, as considered in Section \ref{sec:charact} for $G=G(u_x)$.
}
\hfil

We bring this to the attention of the reader because the implementation in~\cite{GPTC11a} does not explicitly compute $G(u_x)$ 
% MT 08/28/12 - Reviewer wanted it emphasized that we don't actually compute ST(G(u_x)) either.  
nor $ST(G(u_x))$.  
Instead, by the equivalence above, they exist implicitly as the fully-mixed portion of $cl(ST(G))$.  We choose to ignore this technicality in what follows, instead using $ST(G(u_x))$ to refer to the fully-mixed portion of $cl(ST(G))$.  We will take for granted that we have a data-structure encoding of $ST(G(u_x))$ by virtue of the data-structure encoding of $cl(ST(G))$ guaranteed by~\cite{GPTC11a}.  This equivalence will be recalled later as 
% EG 09/23/12 notation modified
%$ST(G(u_x)) \sim cl(ST(G))$.  
$cl(ST(G)) \rightsquigarrow ST(G(u_x))$.  
The advantage of working with $ST(G(u_x))$ is that it allows for the direct application of Theorem~\ref{splitTreeTheorem}.

We apply Theorem~\ref{splitTreeTheorem} as follows.
%Eme-v7 I replaced : with . at the end of the above sentence
%
%\item
 Prior to the node-joins in the second step, we 
 %Marc-v6
 %need to 
 test whether 
 %Marc-v6
 its conditions 
 are satisfied for $ST(G(u_x))$. This can be done one node at a time. 
%Marc-v6
(In case of failure, the graph $G(u_x)+x$ is not a circle graph,  and thus neither is $G+x$.)  More precisely, 
for each node $u$ containing a mixed marker vertex, we test whether $G(u)$ has a chord diagram $C_u$ in which $MP(u)$ is consecutive with mixed marker vertices being bookends.
%Marc-v6
If the test does not fail, then we proceed as follows.  

During 
the contraction step, in addition to a series of node-joins made 
%Marc-v6
by
algorithm \cite{GPTC11a} to contract the fully-mixed subtree, we perform 
%Marc-v6
%Eme-v7 I replace 'the' with 'a', as I wrote before, since the circle joins are not determined a priori (possible choice a priori for every node-join)
%the
a 
corresponding series of 
%Marc-v6. 
circle-joins, just as in the proof of sufficiency of Theorem~\ref{splitTreeTheorem}.
For two nodes $u$ and $v$ to be 
%Marc-v6
joined to form the new node $w$, we need to perform
the circle-join that preserves the consecutiveness and bookends of $MP(u)$ and $MP(v)$ (Lemma~\ref{labelToGraph}). 
 
% \item
 Finally, for the third
 %Marc-v6
step in case 7,
%
% MT 08/28/12 - Reviewer wanted precision: chord isn't inserted, only its endpoints.
the two endpoints representing
 a chord $c$ have to be inserted in the chord diagram $C_{u_x}$ of the node $u_x$ resulting from the contraction step. 
 %Marc-v6:
 This new chord corresponds to $x$ and
 has to cross the chords of $P(u_x)$.
 %Marc-v6
 % and the leaf $x$ corresponds to $c$.
The result is
 a chord diagram for the new prime node labelled with $G(u_x)+x$.
%\end{enumerate}

\end {itemize}

The vertex insertion procedure 
%Marc-v6
for circle graphs that is informally outlined above is captured more precisely as Algorithm~\ref{alg:vertexinsertion}.
The correctness of Algorithm~\ref{alg:vertexinsertion} 
%Marc-v6
follows from
the above discussion, but we prove it more formally in Theorem \ref{th:proof-main-algo} below.
%
%Marc-v6: I found this more confusing than helpful.  In fact, I don't really know what you are referring to.  Is it line 8 in the algorithm?  If so, I think it seems fairly natural and doesn't need comment.  If it's not line 8, then I don't know what you are referring to.  I can't really fix the English because I don't know exactly what you are trying to convey.
%
%Eme-v7 the problem is that we have to maintain not only chord diagrams, but also adjacency lists as done in the split algorithm. But the outline given in Algorithm 2 is written as if both were done simultaneously. For instance at line 5 "insert a chord c..." deals with the chord diagram only, and the next line "add a leaf x..." assumes that the label graph has been updated accordingly. This is a detail but not absolutely correct. Since I thought it would be too heavy to precise each time something like "`and update the adjacencey lists accordingly with algorithm [GPTC]", then I wanted to give a brief explanation before the algorithm. I agree that my explanation is not clear enough. I try a rewriting below, please correct.
%
%%%TO-BE-DONE: read the discussion above and correct the sentence below
%
%Notice that, in order to lighten Algorithm~\ref{alg:vertexinsertion}, we consider that a node, whose label is a circle graph, may be directly labelled by a chord diagram of this circle graph.
%
%Of 
%Marc-v6
%course,
%
%the implementation of this algorithm has to be thought of as complementary to the implementation from \cite{GPTC} where graph labels are encoded as graphs.
%
We point out that the implementation of this algorithm has to be thought of as complementary to the implementation from \cite{GPTC11a}.
%xtof-v8: minor
%Thus, in order to lighten Algorithm~\ref{alg:vertexinsertion}, we consider that a node, whose label is a circle graph, may be directly labelled by a chord diagram of this circle graph,
%though, in the effective implementation using the algorithm \cite{GPTC}, those labels are moreover encoded as graphs (accordingly with those chord diagrams, and using only adjacency lists). 
Thus, in order to lighten Algorithm~\ref{alg:vertexinsertion}, we consider that a node, whose label is a circle graph, may be directly labelled by a chord diagram of this circle graph.
% MT 08/28/12 - Reviewer wanted fewer implementation details from split paper.
%In the effective implementation using the split decomposition algorithm of \cite{GPTC11a}, those labels are in addition encoded as graphs (accordingly with those chord diagrams, and using adjacency lists). 

%Marc-v6: You may want to speak about line 7 in the algorithm in the paragraph above, and the fact that it will be addressed later.  Right now, if someone jumps to read the algorithm, it may throw them off since it hasn't been mentioned at all to this point.

%\begin{algorithm}[ht]
\begin{algorithm}[]
\KwIn{A graph $G$, a vertex $x\notin V(G)$ which is the last vertex in an LBFS ordering of $G+x$, and the split-tree $ST(G)=(T,\mathcal{F})$ equipped with chord diagrams on prime nodes.}
\KwOut{The split-tree $ST(G+x)$ equipped with chord diagrams on prime nodes, if $G+x$ is a circle graph.}
\BlankLine

%\lnl{line:cases} 
\lnl{lnl:beginning}
Determine which case of Theorem~\ref{th:cases} applies 
%Marc-v6: same as before
%thanks to 
based on algorithm \cite{GPTC11a}\;

\medskip
\lnl{lnl:case1-2-4-5-6}
\uIf{case 1, 2, 4, 5 or 6 of Theorem~\ref{th:cases} applies}{
	update $ST(G)$ according to the
	%Marc-v6
	algorithm \cite{GPTC11a},
	 as described in Subsection~\ref{sub:incremental-ST}\;
	}

\medskip	
\lnl{lnl:case3}
\uIf{case 3 of Theorem~\ref{th:cases} applies (let $u_x$ be the unique hybrid prime node)}{
\smallskip
\lnl{lnl:test1}
	\uIf{$MP(u_x)$ is consecutive in the chord diagram $C_{u_x}$ of $u_x$ (Theorem~\ref{splitTreeTheorem})}{
		let $S$ be the factor of the chord diagram $C_{u_x}$ certifying the consecutiveness of $MP(u_x)$\;
		\lnl{lnl:case3-insertion}
		insert in $C_u$ a chord $c$ with endpoints $c_1$ and $c_2$ such that $C_ {u_x}(c_1,c_2)=S$\;
		%update $ST(G)$ according to the algorithm \cite{GPTC} as described in subsection~\ref{sub:incremental-ST}\;
		add a leaf $x$ adjacent to $u_x$ opposite the marker vertex corresponding to chord $c$\;
		}
\smallskip
	\lElse{\Return $G+x$ is not a circle graph\;}		
	}

\medskip	
%Marc-v6
%i
\lnl{lnl:case7}
\uIf{case 7 of Theorem~\ref{th:cases} applies}{
	compute $c\ell(ST(G))$ according to the algorithm \cite{GPTC11a} as described in Subsection~\ref{sub:incremental-ST}\;
	\ForEach{tree-edge $uv$ of the fully-mixed subtree of 
	%Marc-v6: consistency
	$c\ell(ST(G)) \rightsquigarrow ST(G(u_x))$
	the extremities of which are $q_u$ and $q_v$ (they are mixed)
	}{
\lnl{lnl:test-degenerate}
		\uIf{$u$, 
		%Marc-v6: same as before
		(respectively $v$)
		 is degenerate}
		{\uIf{$G(u)$
		%Marc-v6: as above
		(respectively $G(v)$)
		has a chord diagram in which $MP(u)$ is consecutive with mixed marker vertices being bookends}{
	build such a chord diagram $C_u$, respectively $C_v$}
		\lElse{\Return $G+x$ is not a circle graph\;}
		}
\lnl{lnl:test2}
		\uIf{$MP(u)$ is consecutive in $C_u$ with mixed marker vertices being bookends and $MP(v)$ is consecutive in $C_v$ with mixed marker vertices being bookends (Theorem~\ref{splitTreeTheorem})}{
		\lnl{lnl:test2-circle-join}
			perform a circle-join between $C_u$ and $C_v$ with respect to $q_u$ and $q_v$ that preserves consecutiveness and  bookends (Lemma~\ref{labelToGraph})\;
			%perform the node-join between $u$ and $v$ with respect to $q_u$ and $q_v$ as done in the algorithm \cite{GPTC} described in subsection~\ref{sub:incremental-ST}\;
			%consider that $u$ and $v$ are replaced with the above built node, whose graph label has the above built chord diagram\;
			consider that $u$ and $v$ are replaced with a single node whose chord diagram is the above resulting one\;
			}
		\lElse{\Return $G+x$ is not a circle graph\;}
		}
		\smallskip
			%let $u_x$ be the node resulting from the series of node-joins\;
	let $C_{u_x}$ be the chord diagram of the node $u_x$ resulting from the series of circle-joins\;
	let $S$ be the factor of the chord diagram $C_{u_x}$ certifying the consecutiveness of $P(u_x)$\;
\lnl{lnl:case7-insertion}
	insert in $C_{u_x}$ a chord $c$ with endpoints $c_1$ and $c_2$ such that $C_{u_x}(c_1,c_2)=S$\;
%%%%%	%Add a new chord to $C$ defined by the endpoints $\{x_1,x_2\}$, placed so that $MP(u) \cup \{x_1,x_2\}$ is consecutive, and $\{x_1,x_2\}$ are its bookends, following the model in Figure~\ref{fig:addingExample}, and let $C'$ be the result\;
	%%%%%%%update $ST(G)$ according to the algorithm \cite{GPTC} as described in subsection~\ref{sub:incremental-ST}\;
		%add a vertex to $G(u_x)$ with neighborhood $S$ as done in the algorithm \cite{GPTC} described in subsection~\ref{sub:incremental-ST}\;
		add a leaf $x$ adjacent to $u_x$ opposite the marker vertex corresponding to chord $c$\;
	}

\caption{Vertex insertion} \label{alg:vertexinsertion}
\end{algorithm}

\begin{theorem}
\label{th:proof-main-algo}
Given the split-tree $ST(G)$ of a circle graph, equipped with a chord diagram at every prime
%Marc-v6
node,
 and given a good vertex $x$ of $G+x$, Algorithm~\ref{alg:vertexinsertion} tests whether $G+x$ is a circle graph. If so, it returns
 %Marc-v6
$ST(G+x)$, 
equipped with a chord diagram at every prime node.
\end{theorem}
\begin{proof}
Let $\sigma$ be
%Marc-v6
an
LBFS ordering of $G+x$ in which $x$ is good.
The algorithm follows the vertex incremental construction of the split-tree proved in~\cite{GPTC11a}. So if $G+x$ is a circle graph, then the returned GLT is its split-tree $ST(G+x)$. 
% EG 09/23/12 sentence added below
To prove the correctness of the recognition test, we focus on case 7, the other cases are straightforward.
Let us consider the GLT $(T',\mathcal{F'})$ obtained from 
%Marc-v6
$ST(G)$ by contracting
the fully-mixed subtree of $c\ell(ST(G))$ 
%Marc-v6
into
a single node $u_x$ labelled by a graph $G(u_x)$. Observe that $ST(G+x)$ is obtained from $(T',\mathcal{F'})$ by: 
%Marc-v6
(1) 
attaching $x$ as a leaf adjacent to node $u_x$; and 
%Marc-v6
(2)
adding a new marker vertex $q_x$, opposite to leaf 
%Marc-v6
$x$ and 
adjacent to $S=P(u_x)$ in $G(u_x)$. It is proved in~\cite{GPTC11a}, that the resulting node $u'$ and thereby the graph $G(u')=G(u_x)+q_x$ is prime. 

As $G$ is a circle graph, $G(u_x)$ is also a circle 
%Marc-v6
graph, by Corollary~\ref{cor:circle-labels}. 
Likewise, it is clear 
%Marc-v6
from Corollary~\ref{cor:circle-labels}
that $G+x$ is a circle graph if and only if $G(u')$ is a circle graph. 
%Marc-v6
Also,
 by Lemma~\ref{inducedLBFS}, $\sigma_{u'}$ is 
 %Marc-v6
 an 
 LBFS ordering of $G(u')$ and thus $q_x$ is a good vertex of $G(u')$. We apply Theorem~\ref{splitTreeTheorem} to $G(u_x)$ 
 %Marc-v6
 (a circle graph), $G(u')=G(u_x)+q_x$ (a prime graph), and $q_x$ (a good vertex vertex).  By doing so, we conclude
  that $G(u')$ is prime if and only if for every node $v'$ of $ST(G(u_x))$ marked with respect to $S$, $G_{v'}$ has a chord diagram in which $MP(v')$ is consecutive with mixed marker vertices being bookends.
Now observe that, by construction, $ST(G(u_x))$ is isomorphic to the fully-mixed subtree $T_m$ of $c\ell(ST(G))$.
We can thereby conclude that $G+x$ is a circle graph if and only if for every node $v$ of $T_m$, $G(v)$ has a chord diagram in which $MP(v)$ is consecutive with mixed marker vertices being bookends. 
%Marc-v6
Algorithm~\ref{alg:vertexinsertion} 
precisely performs all these tests.

Now assume that $G+x$ is a circle graph. 
%Marc-v6
So as above,
for every node $v$ of the fully-mixed subtree of $c\ell(ST(G))$, there exists a chord diagram $C_v$ in which $MP(v)$ is consecutive with mixed marker vertices being bookends. By Lemma~\ref{labelToGraph}, for every tree-edge $e=vw$ of $T_m$ with extremities $q_v\in V(v)$ and $q_w\in V(w)$, there is a circle-join of $C_v$ and $C_w$ with respect to $q_v$ and $q_w$ that preserves the consecutiveness and bookends. So eventually Algorithm~\ref{alg:vertexinsertion} builds a chord diagram $C_{u_x}$ of node $u_x$ (to which $T_m$ is contracted) such that $MP(u_x)=P(u_x)$ is consecutive. Adding the chord $q_x$, corresponding to the marker vertex opposite 
%Marc-v6
$x$,
 yields a chord diagram of $G(u')$ (which is prime). 
 %Marc-v6
 Therefore
  every prime node of $ST(G+x)$ is equipped with a chord diagram.
\end{proof}

\begin{remark}
At two places in Algorithm~\ref{alg:vertexinsertion} there seem to be possible choices, all leading to 
%Marc-v6
a
 final chord diagram: 
 at line \ref{lnl:test-degenerate} to build a chord diagram of a degenerate node, whose existence (but not unicity) is guaranteed by assumption, 
 and at line \ref{lnl:test2} to perform a circle-join, whose existence (but not unicity) is guaranteed by Lemma~\ref{labelToGraph}.
 In fact, since we obtain a chord diagram of a  prime circle graph, known to be unique up to reflection, we know that, each time, there is a unique possible choice up to reflection.
\end{remark}
%----------------------------------------------------------------------------------------------------------------------
%----------------------------------------------------------------------------------------------------------------------
\section{Data-structure, Implementation and Running Time}
\label{sec:implementation+DS}

The incremental split-tree algorithm from~\cite{GPTC11a} can be implemented as described therein; it runs in time $O(n+m)\alpha(n+m)$. We mention that linear time LBFS implementations appear in~\cite{RTL76} (see also \cite{Gol04}) and~\cite{HMPV00}, and either of these can be used 
%DGC added the see also reference to Golumbic's book.
%Marc-v6
as part of the implementation for~\cite{GPTC11a}.  Thus,
 it remains to implement the routines involved in Algorithm~\ref{alg:vertexinsertion}: consecutiveness test on prime and degenerate nodes;  construction of chord diagrams for degenerate nodes; circle-join 
%Marc-v6
operations
 preserving consecutiveness; and 
 %Marc-v6
 finally,
  chord insertion. To that 
  %Marc-v6
  aim,
   we first describe the data-structure used to maintain a chord diagram at each prime node of the split-tree throughout its construction. We then describe how 
  %Marc-v6: "routine" is outdated
  % the routine mentioned above
Algorithm~\ref{alg:vertexinsertion}
can be implemented in order to obtain the 
%Marc-v6: I would avoid "expected" because of its probability connotations
%expected 
$O(n+m)\alpha(n+m)$ time complexity for the circle graph recognition problem.
%
%The letter $a$ following, resp. preceding, a letter $b$ in a chord diagram is called \emph{successor} or \emph{clockwise-neighbour} of $b$, resp. \emph{predecessor} \emph{counter-clockwise-neighbour} of $b$. Then $a$ and $b$ are called \emph{circle-neighbours}.

%----------------------------------------------------------------------------------------------------------------------
\subsection{Chord Diagram Data-Structure} 
\label{sub:PCCD}

% DGC2 added Emeric's text here.
% MT 08/28/12 - Reviewer mistook the CSC data-structure for circular doubly-linked lists.  I think the reviewer is wrong.  Initially, they are equivalent; but they do not remain so after certain circle join operations.  For example, in lemma 5.6, if we assume the "natural" circular doubly-linked list encoding of each chord diagram, then after the circle join, adjacent pairs of "next" and "prev" pointers will point to each other.  Now the idea of "going in the natural direction" as the reviewer notes doesn't make sense.  It therefore leaves unclear how to recover the chord diagram supposedly being encoded.  To do so, "going in the natural direction" must be defined (we do it via DFS).  And yes, once that is defined, merely reversing the roles of "prev" and "next" will get you the reflection of the encoded chord diagram.  But it still needs to be defined, unlike with a circular doubly-linked list.  They key is that the implicit orientation defined by "next" and "prev" is different in a CSC from a circular doubly-linked list.  However, I do agree with the reviewer's point to the extent that the difference between CSCs and circular doubly-linked list needs to be emphasized.  I think the reviewer asked a natural question.  The difference between them is too subtle to be appreciated without working through the implementation details of each case.  I propose the following introduction that will provide additional context for the eventual definition of CSCs.

We introduce a new data structure for chord diagrams, namely \emph{consistent symmetric cycles}, see below.
At first glance it would seem that
the usual and natural data-structure for chord diagrams would be a circular doubly-linked list; unfortunately, this choice would not 
allow the performance we require.  
In particular, under such a data-structure
each endpoint would be represented by a node with two pointers, say \emph{prev} and \emph{next}, pointing to the endpoint's counter-clockwise and clockwise neighbours, respectively, in the chord diagram.  This would allow consecutive sets of endpoints to be efficiently located and circle-joins to be efficiently performed.  The problem is that our circle graph recognition algorithm sometimes performs circle joins using the reflection of a chord diagram.  In a circular, doubly-linked list, this would require updating all the \emph{prev} pointers to become \emph{next} pointers and vice versa.  That proves too costly.  To achieve the desired running time for circle graph recognition, circle-joins must be performed in constant time.
\medskip

%DGC2 removed the "remark"
%\noindent\emph{Remark.}
%DGC - made the changes requested by Marc so that this paragraph now refers to Figure 6 rather than Lemma 5.6.
One constant-time circle-join alternative using circular, doubly-linked lists would be to simply reinterpret \emph{prev} as \emph{next} and vice versa without actually reassigning pointers.  But this becomes a problem when the circle-join is performed between one chord diagram, say $C$, and the reflection of another, say $C'^r$.  In that case, pairs of \emph{next} pointers will end up pointing to each other and pairs of \emph{prev} pointers will end up pointing to each other.  Figure \ref{fig:reflectionExample} provides one example of a circle-join where this would happen.  In that case, the traditional procedure for traversing a circular, doubly-linked list would no longer work.  Some of the \emph{next} and \emph{prev} pointers need to be interpreted as normal (those from $C$) while the other ones need to be interpreted as the opposite (those from $C'^r$).  The data structure we propose below for chord diagrams generalizes the circular, doubly-linked list to allow for this duality.  
\medskip

%Marc-v6: "For the sake of" isn't the right usage in this context.
%For the sake of time 
%To achieve the desired 
%
%complexity, we need a data-structure for chord diagrams that allows 
%Marc-v6
%us to efficiently 
%
%search a chord diagram and its reflection clockwise and/or counter-clockwise.
% MT 08/28/12 - Adding to strengthen what follows
%In addition
%Marc-v6: Not sure how this can be concluded from the previous sentence.  Confusing at this point.
%Eme-v7 it is to announce the main feature of this data structure. I try a rewriting, please check
%This data-structure will enable us to perform every circle-join in constant time. 
%We shall prove that this data-structure enables to perform every circle-join in constant time. 

% MT 08/31/12 - Reviewer didn't like use of "symmetric" because they claimed it did not support directed pointers such as those intended here.  I don't necessarily agree with their definition of symmetric, or at least with their application of it in this context.  In particular, if there is a pointer a -> b and a pointer b -> a as there is here, then under symmetry that will still hold.  If the reviewer's definition of symmetry doesn't support its use here, then I don't think it supports its use anywhere else in the paper.  I'm not sure how others feel on this.  At the very least, I think the reviewer's suggestion of "opposite arcs" should be avoided for its ambiguity and lack of precision.
\begin{definition}
A \emph{symmetric cycle} is the digraph $C$ obtained from a cycle by replacing every edge 
%Marc-v6
with
 a pair of 
 % EG 09/23/12
 %symmetric arcs. 
 opposite arcs.
 Every vertex $y$ of a symmetric cycle is 
 %Marc-v6: This is a correct use of "thereby".
 thereby associated with two out-neighbours, namely $+_C(y)$ and $-_C(y)$.
%Marc-v6: Same issue as before about separating definitions.  Plus, I think we need to emphasis what is meant by "matched".
\end{definition}

\begin{definition}
Let $C$ be a symmetric cycle on the vertex set $\V = \bigcup_{v \in \V} \{v_1,v_2\}$.  Then each $v_1,v_2$ are said to be \emph{matched}, and $C$ is said to be \emph{consistent} if for every pair $y_1$ and $y_2$ of matched vertices, $+_C(y_1)$ and $+_C(y_2)$ belong to the same connected component of $C-\{y_1,y_2\}$.
\end{definition}

\begin{figure}[htbh]
\begin{center}
\includegraphics[scale=1.05]{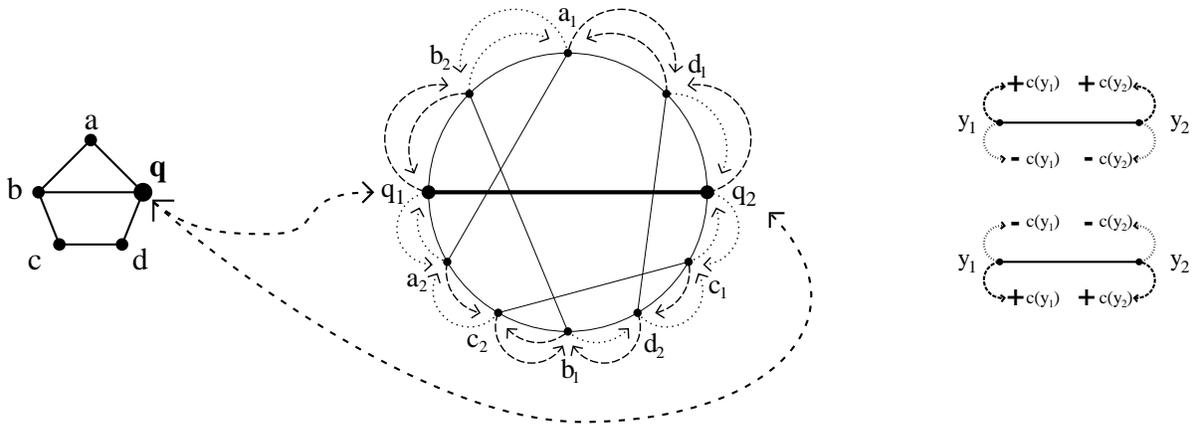}
\end{center}
\vspace{-0.5cm}
\caption{
%Marc-v6:
Example encoding of a circle graph by a CSC.
Arrows represent pointers. The $+/-$ pointer types in the CSC structure are distinguished by two types of dashed arrows. The consistency rule for these pointers is illustrated on the right.}
\label{fig:data-structure}
\end{figure}

Our data-structure for a chord diagram on $V$ implements in the natural way a \emph{consistent symmetric cycle 
%Marc-v6
%(or 
(CSC)} on $\V=\cup_{v\in V}\{v_1,v_2\}$.  That is,
for a chord $y\in V$, the two endpoints $y_1$ and $y_2$ 
% MT 08/28/12 - Adding explicit mention of pointers between endpoints as requested by the reviewer.
are matched with 
% MT 08/31/12 - Despite my comments above about "symmetric" I do think its use here is sufficiently ambiguous to need to be clarified.  I will likewise update other similar uses.  
pointers from each one to the other.  Pointers are also maintained between $y$ and those endpoints.
%symmetric pointers between them.  Symmetric pointers are also maintained between $y$ and those endpoints.
%and symmetric pointers between $y$ and its endpoints are maintained. 
%
%Eme-v7 I add here Marc's remark about reflection (see below), and I complete it with some precison
%
%%%TO-BE-DONE check the precision below, and check if the fact that CSCs encode chord diagrams up to reflection by this way is not a problem in further algorithms (I had not seen this feature, was Christophe aware of it?)
%
Observe that a CSC for chord diagram $C$ is simultaneously a CSC for chord diagram $C^r$. One can distinguish a chord diagram and its reflection by specifying a direction. That is, precisely: chord diagrams up to reflection are encoded by CSCs, and chord diagrams are encoded by CSCs together with the choice of a direction. 
%xtof-v8
%In what follows, we assume that this precision is implied and we will just talk about CSCs as encoding chord diagrams.
In what follows, we assume that this precision is implicit and we will just talk about CSCs as encoding chord diagrams.
%
%Marc-v6: Changing the order of sentences.  Plus, the "on Figure..." mistake was repeated a number of time.  Here, "illustrated" needed two "l"'s.
This data-structure is illustrated in Figure \ref{fig:data-structure}.
Observe that searching a CSC 
%Eme-V7 now I remove the following precision, contradicting the above one
%(representing a chord diagram) 
in a given direction is achieved in linear time by a depth-first search (DFS).  

Let $y_1$ and $y_2$ be the endpoints of chord $y$ of the chord diagram $C$. Let $+_C(y_1,y_2)$ denote the sequence of endpoints (other than $y_1$ and $y_2$)
%Marc-v6
%, 
encountered while starting a DFS on $C$ from $y_1$ with pointer $+_C(y_1)$ and stopping at $+_C(y_2)$. The sequences $-_C(y_1,y_2)$, $+_C(y_2,y_1)$ and $-_C(y_2,y_1)$ are defined similarly. Observe that the sequence $+_C(y_1,y_2)$ is the reversal of $+_C(y_2,y_1)$.
The following observation establishes the links between the CSC representation of a chord diagram $C$ and its representation by a circular word.

\begin{observation} \label{obs:consistent-PCCD}
If $y_1$ and $y_2$ are the endpoints of chord $y$ in a chord diagram $C$, then 
%Marc-v6
%either 1 or 2 holds:
exactly one of the following holds:
\begin{enumerate}
\item $+_C(y_1,y_2)=C(y_1,y_2)$ 
%Marc-v6
(and thus $+_C(y_2,y_1)=C^r(y_1,y_2)$, $-_C(y_1,y_2)=C^r(y_2,y_1)$ and $-_C(y_2,y_1)=C(y_2,y_1)$)
\item $+_C(y_1,y_2)=C^r(y_2,y_1)$ 
%Marc-v6
(and thus $+_C(y_2,y_1)=C(y_2,y_1)$, $-_C(y_1,y_2)=C(y_1,y_2)$ and $-_C(y_2,y_1)=C^r(y_1,y_2)$)
\end{enumerate}
\end{observation}

%Marc-v6: Isn't this part below just repeated from the beginning of section 4?  Yes, the stuff below is slightly more detailed, but it doesn't seem to be used or required anywhere else in the paper.  That is, are any of the details here used elsewhere in the paper?  Seems like an artifact of previous drafts.
%
%Eme-v7: it is not an artifact from previous drafts, it is something that Christophe added especially. He wanted the reader to have an overview of the data structure from [GPTC], somehow precise but not too much. I felt the same as you, and I would agree to remove this part which in my opinion makes things appear more complicated and brings not much (I think that the only important thing from [GPTC] is that states are recorded, but this is said at the beginning of section 5.2 and seems enough to me). But let us discuss it with other coauthors. I mark it as to-be-done... Anyway, I think we should write one sentence to say that a precise data structure is given in [GPTC], I add one below
%
%%%TO-BE-DONE: read the above discussion, and decide if the paragraph removed below was worth being in the paper (end of subsection 5.1 in v5 draft), or if the new sentence below is enough
%
For the sake of implementation, this data-structure for circle graphs completes the data-structure used
%DGC omitted: " and described precisely" 
in~\cite{GPTC11a} to represent 
%a GLT $(T,\mathcal{F})$. 
%(and thus 
the split-tree $ST(G)$ of a graph $G$.
\subsection{Implementation with CSCs}

%Marc-v6
This section uses CSCs to implement
the routines involved in Algorithm \ref{alg:vertexinsertion} and 
%Marc-v6
evaluates 
their costs. The computation of the perfect/empty/mixed states 
%Marc-v6: clarify
for marker vertices of nodes is handled 
in the algorithm from \cite{GPTC11a} (here at line \ref{lnl:beginning} in Algorithm \ref{alg:vertexinsertion}). Notably, the set of non-empty marker vertices $MP(u)$ is 
%Marc-v6
also computed by~\cite{GPTC11a}, and
assumed to be known for every involved node $u$.
%Marc-v6: Thought it was important enough to mention again.
It is also important to remind the reader that a CSC simultaneously encodes a chord diagram $C$ and its reflection $C^r$.  This will be crucial for the efficiency of the implementation of Algorithm~\ref{alg:vertexinsertion}.
%

%----------------------------------------------------------------------------------------------------------------------
\subsubsection{Testing Consecutiveness in a CSC}

% MT 08/28/12 - Reviewer noted that a circle join doesn't really apply in line 7 of the algorithm.  Replacing it with something more direct.
There are three different times during algorithm~\ref{alg:vertexinsertion} when 
%A node $u$ of the fully-mixed subtree of 
%Marc-v6: I would suggest getting rid of the parentheses.  This would be more aesthetically pleasing and reinforce that important paragraph from before.  I haven't changed this, but have left it as a recommendation easily changed by a "find and replace all".  
%Eme-v7 agreed
%$c\ell(ST(G))$ $\rightsquigarrow ST(G(u_x))$ may be involved in 
%Marc-v6
%a
% 
%circle-join. Prior to such a circle-join, 
we need to test whether $G(u)$ has a chord diagram $C_u$ in which the chords of $MP(u)$ are consecutive with mixed marker vertices being bookends (lines \ref{lnl:test1}, \ref{lnl:test-degenerate} and \ref{lnl:test2}).  We also have to build such a chord diagram 
if the node is degenerate (line \ref{lnl:test-degenerate} in Algorithm~\ref{alg:vertexinsertion}).  Recall that a prime label is already equipped with its chord diagram. 
%Eme-v7 I change the sentence below, since in the non-degenerate case we do just a test, not a building
%We argue 
%below that the chord diagrams for degenerate nodes can be built
%in constant time if $u$ is a degenerate node and in $O(|MP(u)|)$ otherwise. 
We argue 
below that, if $u$ is a degenerate node, 
then this test can be performed (and a chord diagram can be built) in constant time; 
  and otherwise, this test can be performed in $O(|MP(u)|)$.
\smallskip

%Marc-v6: This relates to my earlier issue with proposition 3.10.  Given the preceding sentence, I don't think we need the next sentence.  If anything, it only seems to be here to recall that (by now long forgotten result).  If you insist on including proposition 3.10, then maybe include it here.
%
%Eme-v7 see comments about Theorem 3.10 (ex-Prop 3.10) which was crucial. This theorem is convenient for the next Lemma, hance I put back the next sentence.
%
The case of degenerate nodes follows directly from Theorem \ref{prop:ST-prime}, satisfied by $ST(G(u_x))$.
%

%The case of degenerate nodes follows from Lemma~\ref{mixed} and the following property of $c\ell(ST(G))$ observed in~\cite{GPTC}:
%
%\begin{remark} ~\label{rem:2-non-mixed}
%Let $ST(G)$ be marked with respect to $N(x)$ with $x$ a good vertex of $G+x$. If $u$ is a degenerate node of the fully-mixed subtree of $c\ell(ST(G))$, then $V(u)$ contains at most one perfect marker vertex and at most one empty marker vertex.
%\end{remark}

%Marc-v6: Again, the proposition 3.10 issue.  The next lemma's proof is the only other place where it's used.  I don't think it's necessary.  Recall that earlier drafts did not use anything similar.  The conditions for a degenerate node having the consecutive property can be simply stated as in those drafts, and they are obvious enough from the form of their chord diagrams to not need a proof.  If we go that route, then we can avoid proposition 3.10 altogether along with its issues.  Nevertheless, I have corrected the English below in case you wish to keep it.
%
%Eme-v7 Theorem 3.10 gives a direct answer, so let us use it. If we do not know that the size of degenerate nodes is bounded, then we need to describe precisely every possible case to check that these cases can be tested in constant time... older drafts were correct but not explicit enough on this point.
%
\begin{lemma} \label{lem:degen-consecutive-test}
Let $ST(G)$ be marked with respect to $N(x)$ with $x$ a good vertex of $G+x$. If $u$ is a degenerate node of the fully-mixed subtree of $c\ell(ST(G))$ $\rightsquigarrow ST(G(u_x))$, then testing if there exists a CSC for a chord diagram $C_u$ of $G(u)$ in which $MP(u)$ is consecutive with mixed marker vertices being bookends,
%Marc-v6
and, 
%
% MT 08/31/12 - Reviewer wanted all O(1) replaced by "constant time".  I'm only commenting this one.
if so computing it, requires constant time.
\end{lemma}

\begin{proof}
Assume there exists a chord diagram $C_u$ in which $MP(u)$ is consecutive with mixed marker vertices being bookends. 
%Marc-v6
Therefore $MP(u)$ contains at most two mixed marker vertices.  Applying Theorem \ref{prop:ST-prime}
to $ST(G(u_x))$, we see that $u$ has at most two non-mixed marker vertices.  
%Marc-v6
Hence,
$u$ contains at most four marker vertices. The number of possible chord diagrams is thereby bounded by a constant (there are at most 8 chord endpoints to arrange).
%
%It follows that if $u$ is a clique node, then a chord diagram $C_u$ is of the form $C_u=A.A$ where $A=\pi(V(u))$ is any permutation of $V(u)$. If $u$ is a star node with centre marker vertex $q$, a chord diagram $C_u$ is of the form $C_u=q_1.A.q_2. A^r$ where $A=\pi(V(u)\setminus\{q\})$ is any permutation of $V(u)\setminus\{q\}$ and  $A^r$ its reversal. 
%
%Marc-v6
Thus, 
the construction of an appropriate chord diagram $C_u$, or the test that no appropriate chord diagram exists, can be done in constant time.
Let us recall that chord diagrams of degenerate nodes have the form demonstrated in Figure~\ref{fig:degenerateExample}. 
\end{proof}

We now consider the case of a prime node $u$. 
%Marc-v6: "Keep in mind" too colloquial.
%Keep in mind 
%from the split decomposition algorithm, 
%
%
%Recall
%
%that the set $MP(u)$ is given, as well as the 
%Marc-v6
%Eme-v7 it was really (unique up to reflection)  since this has no importance here, G(u) is just given with A chord diagram C_u. I try a little change below
%unique (up to reflection) 
%
%chord diagram $C_u$ of $G(u)$.
%
Recall that the set $MP(u)$ is given, as well as a 
chord diagram $C_u$ of $G(u)$ (in fact unique up to reflection).

\begin{lemma} \label{lem:prime-consecutive-test}
Let $ST(G)$ be marked with respect to $N(x)$ with $x$ a good vertex of $G+x$. If $u$ is a prime node, %Marc-v6
then
testing if $MP(u)$ is consecutive with mixed marker vertices being bookends in the chord diagram $C_u$ of $G(u)$ requires $O(|MP(u)|)$ time.
\end{lemma}
\begin{proof}
% MT 08/28/12 - Reviewer didn't like the way the proof was structured by first assuming the very thing we are meant to determine.
Recall that $MP(u)$ can be assumed to have been computed by the split algorithm of~\cite{GPTC11a}.  Consider some marker vertex $q \in MP(u)$.  If $S_e$ is a set of endpoints certifying that $MP(u)$ is consecutive in $C_u$, then $q_1 \in S_e$ or $q_2 \in S_e$ but not both.  Moreover, $S_e$ is of the form
%Assume 
%Marc-v6
%that
%
%$MP(u)$ is consecutive and let $S$ be the set of endpoints 
%(factor of $C_u$)
% certifying the consecutiveness.
%Then every marker vertex $q\in MP(u)$ has exactly one endpoint, say $q_1$, in $S$. 
%Marc-v6
%Moreover,
%
 %$S$ is of the form 
%Marc-v6: I had trouble with this.  It looks like this is some kind of notation rather than a label.  I started looking for its definition somewhere earlier in the paper.  I don't think it's as natural as you had intended it.  I suggest finding some other label for these.
%
%Eme-v7 S is in fact a word, not a set, so I add "(factor of $C_u$)" above... and I also added a precision in the circle graph prelimiaries to say that we may make this abuse set<->factor
$S_e^- q_1 S_e^+$ or 
% MT 08/28/12 - Adding one for q_2 to be more precise and to reflect new proof structure. 
$S_e^-q_2S_e^+$ with $S_e^-$ and $S_e^+$ 
being possibly empty words. So, to test the consecutiveness of $MP(u)$, it suffices to test the existence of these sets $S_e^-$ and $S_e^+$ of endpoints. To that
%Marc-v6
aim,
 proceed as follows: search $C_u$ from $q_1$ in one direction, say using the pointer $-_{C_u}(q_1)$, as long as the encountered endpoint corresponds to a marker vertex $q'$ of $MP(u)$ and the other endpoint of $q'$ has not yet been discovered.  
 % MT 08/28/12 - Reviewer clarified need to have bi-directional pointers between each chord's endpoints to effectively perform this test in constant time.
 Using the pointers between the endpoints of each chord, it can be determined in constant time if the other endpoint has already been discovered.
 Perform the same search in the other direction, i.e. with the pointer $+_{C_u}(q_1)$. 
% MT 08/28/12 - Adding one for q_2 to be more precise and to reflect new proof structure.
If $S_e^- q_1 S_e^+$ isn't located in this, then perform the same search, but this time starting at $q_2$.
With these searches, the existence of $S_e$ can be determined
%
% Clearly, the two searches can be 
 %Marc-v6
% performed
 %
  in $O(|MP(u)|)$ time with DFS. Once this test has been performed, testing if  non-bookend elements of $S$ are non-mixed has the same cost $O(|MP(u)|)$. 
\end{proof}

%----------------------------------------------------------------------------------------------------------------------
\subsubsection{Circle-Joins Preserving Consecutiveness (with CSCs)}

%Marc-v6: I had trouble with this paragraph.  The first sentence is awkward English making it difficult to understand.  The second sentence is too important to be covered so briefly, especially considering that to this point in the paper we have not mentioned anything about the CSC's being reflection-invariant (see my earlier comment and addition to the text).  You are asking a lot of the reader to make that connection.  In addition to my earlier addition regarding reflection-invariance, I suggest the rewrite that follows.
%We want to prove that in the condition of Lemma~\ref{labelToGraph}, we can identify in constant time, amongst the four possible the circle-joins, the one that preserves consecutiveness and bookends (line \ref{lnl:test2-circle-join} in Algorithm~\ref{alg:vertexinsertion}). The use of symmetric cycle is important here as we may use the reflection of a chord diagram and we want to avoid the search of the whole chord diagram in order to reassign direction pointers.
%
%Eme-v7 yes ! much more clear and interesting
%
%DGC minor wording change below
We want to prove that we can identify -- in constant time -- which of the four possible circle-joins of Lemma~\ref{labelToGraph} preserves consecutiveness and bookends (line~\ref{lnl:test2-circle-join} in Algorithm~\ref{alg:vertexinsertion}).  Recall that some of the constructions from Lemma~\ref{labelToGraph} use the reflection of the chord diagram.  Our use of consistent symmetric cycles and their property of being invariant under reflection (Section~\ref{sub:PCCD}) is important in this regard: it means that no additional work is required to compute the reflection of a chord diagram in implementing the circle-joins of Lemma~\ref{labelToGraph}.

\begin{lemma} \label{PCCDtime}
Let $C_u$ and $C_v$ be two chord 
%Marc-v6
diagrams, respectively,
 on the set of chords $V(u)$ and $V(v)$. Let $S_u\subseteq V(u)$ and $S_v\subseteq V(v)$ be consecutive 
 %Marc-v6
 sets
  of chords 
  %Marc-v6
  in $C_u$ and $C_v$, respectively.
Given the CSCs for $C_u$ and $C_v$, the bookends $q$ and $q'$ of 
%Marc-v6
$S_u$,
 and the bookends $r$ and $r'$ of $S_v$, one can build in constant time a CSC for a chord diagram $C$ on $(V(u)\setminus\{q\})\cup (V(v)\setminus\{r\})$ satisfying the conclusion of Lemma~\ref{labelToGraph}, which we recall 
 %Marc-v6
 %is that:
 as
%PCCDs for the chord diagrams $C_u$ and $C_v$ respectively on the set of chords $V(u)$ and $V(v)$. Let $S_u\subseteq V(u)$ and $S_v\subseteq V(v)$ be consecutive set of chords in their respective PCCD such that $q$ is the unique bookend of $S_u$ and $r$ the unique bookend of $S_v$. Then in $O(1)$ time, one can build a PCCD for the chord diagram $C$ on $(V(u)\setminus\{q\})\cup (V(v)\setminus\{r\})$ such that:
\begin{enumerate}
\item $C$ results from a circle-join $\odot$ or $\hat\odot$ of $C_u$ and $C_v$ or $C^r_v$ with respect to $q$ and $r$, 
\vspace{-0.2cm}
\item $S=(S_u\setminus\{q\})\cup (S_v\setminus\{r\})$ is consecutive in $C$,
\vspace{-0.2cm}
\item $S$ has bookends $q'$ and $r'$.
\end{enumerate}
\end{lemma}                                                                                                                                                                                                                                                                                                                                                                                                                                                                      
\begin{proof}
%Marc-v6: Same issue as before with w.o.l.o.g.
%Without loss of generality assume the following: 
We will address the following case 
%Eme-v7 similar better than symmetric, as Marc noticed before
%(the others are symmetric):
(the others are similar):
$C_u(q_1,q'_1)$ and $C_v(r_1,r'_1)$ respectively
certify the consecutiveness of $S_u$ in $C_u$ and of $S_v$ in $C_v$; $C_u(q_1,q'_1)$ and $C_v(r_1,r'_1)$ 
%Marc-v6
are, respectively,
strictly contained in $C_u(q_1,q_2)$ and $C_v(r_1,r_2)$; $C_u(q_1,q_2)=+_{C_u}(q_1,q_2)$ and $C_v(r_1,r_2)=+_{C_v}(r_1,r_2)$. 
%Marc-v6
%The other cases are symmetric. 
By Observation~\ref{obs:consistent-PCCD}, we have

\smallskip
%Marc-v6: Again, I didn't love the use of "." here for concatenation.  See my earlier comment on this.  It just doesn't play nicely with the other symbols below.  Can we find something better?
\centerline{$(C_u,q)\odot (C_v,r)\sim  +_{C_u}(q_1,q_2) +_{C_v}(r_1,r_2) -_{C_u}(q_2,q_1) -_{C_v}(r_2,r_1)$}
\smallskip
\centerline{$(C_u,q)~\hat\odot~ (C_v,r)\sim  +_{C_u}(q_1,q_2) -_{C_v}(r_2,r_1) -_{C_u}(q_2,q_1) +_{C_v}(r_1,r_2)$}

\smallskip
\centerline{$(C_u,q)\odot (C^r_v,r)\sim  +_{C_u}(q_1,q_2) -_{C_v}(r_1,r_2) -_{C_u}(q_2,q_1) +_{C_v}(r_2,r_1)$}

\smallskip
\centerline{$(C_u,q)~\hat\odot~ (C^r_v,r)\sim  +_{C_u}(q_1,q_2) +_{C_v}(r_2,r_1) -_{C_u}(q_2,q_1) -_{C_v}(r_1,r_2)$}

\medskip
By Lemma~\ref{labelToGraph}, one of the 
%Marc-v6
four chord diagrams above
preserves consecutiveness and bookends. 
%Marc-v6
Under
 the assumptions above, $S$ is consecutive
%Marc-v6: similar sentence construction issue with things being too left too late or broken up by subordinate clauses.  Doesn't work so well in English anymore
in $C=(C_u,q)~\hat\odot~ (C^r_v,r)$, with bookends $r'$ and $q'$.
The CSC for $C$ is obtained from those for $C_u$ and $C_v$ by reassigning a constant  number of pointers. 
%Marc-v6: I found this confusing.
%
%Eme-v7 this proof with much notations and subscripts would need to be precisely checked, I did not do it and I do not do it now. Did Marc do it or was this part of the confusion? Anyway, someone should do it, in v5 it was Christophe's first writing of it.
%%%TO-BE-DONE check the proof as commented above
%
%Under our assumptions (the other cases are symmetric), if we have:
For example, assuming the following 
%Eme-v7 similar
%(the other cases are symmetric):
(the other cases are similar):
\begin{center}
$-_{C_u}(q_1)=a_u$ and $+_{C_u}(a_u)=q_1$;
$+_{C_u}(q_1)=b_u$ and $-_{C_u}(b_u)=q_1$;\\
$+_{C_u}(q_2)=c_u$ and $+_{C_u}(c_u)=q_2$;
$-_{C_u}(q_2)=d_u$ and $+_{C_u}(d_u)=q_2$;\\
$-_{C_v}(r_1)=a_v$ and $+_{C_v}(a_v)=r_1$;
$-_{C_v}(r_1)=b_v$ and $-_{C_v}(b_v)=r_1$;\\
$-_{C_v}(r_2)=c_v$ and $+_{C_v}(c_v)=r_2$;
$-_{C_v}(r_2)=d_v$ and $+_{C_v}(d_v)=r_2$;\\
\end{center}
then we perform the following updates:\\
\centerline{$+_C(b_v)=b_u$ and $-_C(b_u)=b_v$; $+_C(c_u)=a_v$ and $+_C(a_v)=c_u$;}
\centerline{$+_C(d_v)=d_u$ and $+_C(d_u)=d_v$; $+_C(a_u)=c_v$ and $+_C(c_v)=a_u$.}

\medskip
It is not difficult to check that the above pointer reassignments preserve the consistency property.  
% MT 08/28/12 - Adding an extra detail to directly address the "bug" noted by the reviewer in which pointers between endpoints are needed for this to be constant time.
Regarding the running time, observe that only a constant number of pointer reassignments are required.  Moreover, given the bookends $q_1,q_1',r_1,r_1'$, their other endpoints $q_2,q_2',r_2,r_2'$, respectively, can be accessed in constant time using the pointers between the endpoints of each chord.  And to decide
in constant time which of the possible circle-joins we need to 
%Marc-v6
perform,
it suffices to store a constant size table describing every possible
%Marc-v6
case, along with the required circle-join operation for that case.
\end{proof}

%\begin{lemma} ~\label{lem:contraction}
%Let $ST(G)$ be marked with respect to $N(x)$ with $x$ a good vertex of $G+x$. Then computing the CSC of the node resulting from the contraction of $c\ell(ST(G))$ by a series of circle-join requires $O(|MP(x)|)$.
%\end{lemma}
%\begin{proof}
%By definition of the fully-mixed subtree, the number $c\ell(ST(G)))$, 

%%The result partly follows from the fact that the number $t$ of tree-edges in $c\ell(ST(G))$, and so the number of circle-joins to perform, is bounded by $O(|N(x)|)$. Indeed $t$ is the number of tree-edges of the fully-mixed subtree of $ST(G)$, which by~\cite{GPTC}, is 

%

%\end{proof}

%----------------------------------------------------------------------------------------------------------------------
\subsubsection{Chord Insertion in a CSC} \label{sec:primeChordDiagram}

To complete the implementation of Algorithm~\ref{alg:vertexinsertion}, it remains to describe how a new chord $c$ can be inserted in a CSC. 
This task occurs at lines~\ref{lnl:case3-insertion} and~\ref{lnl:case7-insertion}.
%This task occurs at line~\ref{lnl:case3-insertion}, in the CSC of an existent prime node, and at line~\ref{lnl:case7-insertion}, in the CSC of the node resulting from the contraction of the fully-mixed subtree of $c\ell(ST(G))$ by a series of circle-join. 
In both cases the resulting chord diagram $C$ corresponds to a prime graph. Moreover, thanks to the previous steps, the neighbourhood of the vertex represented by $c$ is consecutive in $C$.

\begin{lemma} ~\label{lem:chord-insertion}
Given a CSC for a chord diagram $C$ and the bookends of a consecutive set $S_e$ of endpoints in $C$, the insertion of a new chord $c$ intersecting exactly the chords with an endpoint in $S_e$ requires constant time.
\end{lemma}

\begin{proof}
Let $b$ and $b'$ be the bookends of $S_e$ and let $a\notin S_e$ and $a'\notin S_e$ be the endpoints neighbouring 
%Marc-v6
$b$ and $b'$, respectively. 
It suffices to reassign a constant number of pointers towards the endpoints $c_1$ and $c_2$ of the new chord
%Marc-v6
$c$.  For example, if $+_C(a)=b$ and $+_C(b)=a$, then set $+_C(a)=c_1$ and $+_C(b)=c_1$; and if $-_C(b')=a$ and $+_C(a')=b'$, then set $-_C(b')=c_2$ and $+_C(a')=c_2$.  The other cases are symmetric.  Following that,
we need to initialize the pointers of $c_1$ and $c_2$ in a consistent way: for example $+_C(c_1)=b$ and $+_C(c_2)=b'$.
%To insert the new chord $c_x$ representing $x$, it suffices to identify the set $S_e$ of endpoints of $C$ certifying the consecutiveness of $N(x)$ and then insert $c_x$ so that its extremities $x_1$ and $x_2$  becomes the bookends of $S_e\cup\{x_1,x_2\}$. Observe that $S_e$ can be identified in time $O(|N(x)|)$: pick and arbitrary vertex $y\in N(x)$ and perform a search from its of each endpoints in each directions until an endpoint of a vertex not in $N(x)$ is found. As $N(x)$ is consecutive, this process eventually find $S_e$ and clearly runs in $O(|N(x)|)$. Finally, once $S_e$ is known, insert $x_1$ and $x_2$ in a consistent way: for example $+_C(x_1)$ and $+_C(x_2)$ point towards elements of $S_e$. This requires only requires a cosntant number of pointers updates and initializations.
\end{proof}

%----------------------------------------------------------------------------------------------------------------------
\subsection{The Running Time}

%Marc-v6: I found the following paragraph redundant.  
%In~\cite{GPTC}, the authors designed an algorithm that uses and maintains the data-structure described above (not equipped with the CSC's on prime nodes) encoding  the split-tree $ST(G)$ of a graph $G$.  That algorithm is a LBFS vertex incremental algorithm.
%in time $O(\alpha(n+m)(n+m))$, where $\alpha$ is the inverse of Ackermann's function. 
%Eme-v7 I agree with all these modifications

%Marc-v6: Move the paragraph following the theorem up here as an introduction instead of the paragraph above.
As already 
%Marc-v6
described,
compared to the LBFS incremental split decomposition algorithm
%Marc-v6
of~\cite{GPTC11a}, 
the circle graph recognition problem 
%Marc-v6: Doesn't work in English.
%requires to 
must only
handle the consecutiveness test %(line~\ref{lnl:test1} and~\ref{lnl:test2} 
of 
%Marc-v6
Algorithm~\ref{alg:vertexinsertion}
and the maintenance of CSCs for the chord diagrams of prime nodes. So if we prove that, at each vertex insertion, these tasks can be performed in time linear in the cost of the split-tree modifications, then we could conclude that the circle graph recognition problem can be solved 
%Marc-v6: Awkward in English
%in  the same complexity than 
as efficiently as the split decomposition algorithm.  
%Marc-v6: New bridge to the theorem.
Regarding the latter,~\cite{GPTC11a} proved the following:

\begin{theorem} [Theorem 6.21 in~\cite{GPTC11a}]\label{th:ST-algo}
The split-tree $ST(G)$ of a graph $G=(V,E)$ with $n$ vertices and $m$ edges can be built incrementally according to an LBFS ordering in time  $O(n+m)\alpha(n+m)$, where $\alpha$ is the inverse Ackermann function.
\end{theorem}

%As already mentioned, compared to the LBFS incremental split decomposition algorithm, the circle graph recognition problem requires to handle the consecutiveness test %(line~\ref{lnl:test1} and~\ref{lnl:test2} 
%of Algorithm~\ref{alg:vertexinsertion}) and the maintenance of CSC's for the chord diagrams of prime nodes. So if we prove that, at each vertex insertion, these tasks can be performed in time linear in the cost of the split-tree modifications, then we could conclude that the circle graph recognition problem can be solved in  the same complexity than the split decomposition algorithm.

For
%Marc-v6
an
LBFS ordering  $\sigma=x_1<\dots <x_n$ of a graph $G$, 
%and $G_i$ be the subgraph of $G$ induced by $V_i=\{x\mid \sigma(x)\leqslant i\}$.
let $G_i$ be the subgraph of $G$ induced by $V_i=\{x_1,\dots, x_i\}$. 
Let \texttt{insertion-cost}$(x_i,ST(G_{i-1}))$ denote the complexity 
%Marc-v6
%cost spent by 
of
the LBFS incremental split decomposition algorithm \cite{GPTC11a} to compute $ST(G_i)$ from $ST(G_{i-1})$ marked with $N_i(x_i)=N(x_i)\cap V_{i-1}$.
From Theorem~\ref{th:ST-algo} we 
%Marc-v6
have:

$$\sum_{i=1}^n \mbox{\texttt{insertion-cost}}(x_i,ST(G_{i-1}))~\in~ O(n+m)\alpha(n+m)$$

\begin{theorem}
The circle graph recognition test can be performed in time $O(n+m)\alpha(n+m)$ on any graph on $n$ vertices and $m$ edges.
\end{theorem}
\begin{proof}
Let $\sigma=x_1<\dots <x_n$ be 
%Marc-v6
an
 LBFS ordering of the graph $G$. Assume that $ST(G_{i-1})$, marked with respect to $N_i(x_i)$, is equipped with a CSC at every prime node. We prove that computing $ST(G_i)$ and the CSCs of its prime nodes (if $G_i$ is a circle graph) requires 
$O($\texttt{insertion-cost}$ (x_i,$ $ST(G_{i-1})))$.

%Marc-v6
First,
 observe that in cases 1, 2, 4, 5, and 6 of Theorem~\ref{th:cases} (line \ref{lnl:case1-2-4-5-6} in Algorithm~\ref{alg:vertexinsertion}) the prime nodes of $ST(G_{i-1})$ are not affected by 
 %Marc-v6
 $x_i$'s
 insertion. 
 %Marc-v6
 So none of the CSCs stored at prime nodes are affected,
and thus no extra work is required for the circle graph recognition problem.

%Marc-v6
Now assume that 
case 3 of Theorem~\ref{th:cases} holds (line \ref{lnl:case3} in Algorithm~\ref{alg:vertexinsertion}). Let $u_x$ denote the unique prime hybrid node of $ST(G_{i-1})$. We need to insert a chord $c$ in the CSC for the chord diagram $C_{u_x}$ of $G(u_x)$ which 
%Marc-v6
exactly intersects
the chords in $MP(u_x)$. As $u_x$ is the only prime node affected by $x_i$'s insertion
%Marc-v6
(Theorem~\ref{splitTreeTheorem}), 
$G_i$ is a circle graph if and only if $MP(u_x)$ is consecutive in $C_{u_x}$. As $ST(G_{i-1})$ is marked with respect to $N_i(x_i)$ (i.e. $MP(u_x)$ is identified by the split-tree algorithm), 
%Marc-v6: move later.
%by Lemma~\ref{lem:prime-consecutive-test} 
testing the consecutiveness of $MP(u_x)$ requires $O(|MP(u_x)|)$ time,
%Marc-v6
by Lemma~\ref{lem:prime-consecutive-test}.  
Moreover by Lemma~\ref{lem:chord-insertion}, inserting the chord $c$ only takes constant time. The total amount of time spent to update the CSC for $C_{u_x}$ is clearly 
%Marc-v6
$O($\texttt{insertion-cost}$(x_i,ST(G_{i-1})))$,
since $MP(u_x)$
%Marc-v6
has
been computed by the split decomposition tree algorithm (at this step $i$).

%Marc-v6
Finally, assume
that case 7 of Theorem~\ref{th:cases} holds (line \ref{lnl:case7} in Algorithm~\ref{alg:vertexinsertion}). Let $u_x$ denote the node resulting from the contraction of the fully-mixed subtree $T_m$ of $c\ell(ST(G_{i-1}))$ $\rightsquigarrow ST(G(u_x))$. We need to compute a CSC for the chord diagram $C_{u_x}$ of $G({u_x})$ such that $MP({u_x})$ is consecutive and then insert a new chord, say $c_x$, 
%Marc-v6
%inserting 
exactly intersecting
 $MP({u_x})$. Again, by Theorem~\ref{splitTreeTheorem}, this is possible (i.e. $G_i$ is a circle graph) if and only if every node $v$ of $T_m$ has a chord diagram in which $MP(v)$ is consecutive with mixed marker vertices being bookends. This property of $MP(v)$ can be tested and built in constant time if $v$ is a degenerate node of $c\ell(ST(G_{i-1}))$ (Lemma~\ref{lem:degen-consecutive-test}), and can be tested in $O(|MP(v)|)$ time if $v$ is a prime node of $T_m$ (by Lemma~\ref{lem:prime-consecutive-test}).
The sum of these costs over involved nodes $v$ is $O($\texttt{insertion-cost}$(x_i,ST(G_{i-1})))$ since $MP(v)$ is computed by the split decomposition algorithm \cite{GPTC11a} for each $v$.
 Moreover, by Lemma~\ref{PCCDtime}, with a constant time extra cost, a circle-join preserving consecutiveness and bookends can be performed in parallel to every node-join operation  required to contract $T_m$ into ${u_x}$ 
 %Marc-v6
 that is performed by the split decomposition algorithm \cite{GPTC11a},
 with total cost $O($\texttt{insertion-cost}$(x_i,ST(G_{i-1})))$. Finally as in the previous case, we eventually insert the new chord $c_x$ in the CSC for the resulting chord diagram $C_{u_x}$. By Lemma~\ref{lem:chord-insertion}, this also requires constant time since $MP({u_x})$ is known. 
 %Marc-v6: I used "finally" to begin the paragraph, can't use it again to end it.
 %Finally, the total 
  In total, the
  amount of time spent to built the CSC of the new prime node is  $O($\texttt{insertion-cost}$(x_i,ST(G_{i-1})))$.
\end{proof}

\section{Concluding Remarks}

% MT 08/31/12 - Reviewer wanted a conclusion.  It directly addresses the reviewer's questions about the utility of knowing the split decomposition a priori and possible future applications to rank-width.  I did not think the other suggestions for the conclusion were especially relevant given the direction of the paper.
This paper presents the first subquadratic circle graph recognition algorithm.  It also develops a new characterization of circle graphs in terms of LBFS (upon which the algorithm is based).  The algorithm operates incrementally, extending the incremental split decomposition algorithm from the companion paper~\cite{GPTC11a}.  The two operate in parallel.  As each new vertex is inserted, the circle graph recognition algorithm inspects properties of the split-tree to determine if the resulting graph will remain a circle graph.  If it does, the split-tree is updated to account for the new vertex.  The running time for the entire process is $O(n+m)\alpha(n+m)$, where $\alpha$ is the 
% EG 09/23/12 - NOT inverse of Ackermann's function, 
inverse Ackermann function, 
which is essentially constant for all practical graphs.  It is important to note that this $\alpha$ factor is due to the split decomposition algorithm; the circle portion is consistent with linear time.  Thus, a linear time implementation of the split decomposition portion would result in a linear time circle graph recognition algorithm.  

Eliminating the dependence on the incremental split decomposition portion may prove difficult.  Recall that split decomposition reduces the problem of recognizing circle graphs to that of recognizing prime circle graphs.  But since prime graphs cannot be further decomposed, simply knowing the split decomposition a priori does not help.  Therefore bypassing the incremental split decomposition portion above may necessarily mean bypassing split decomposition altogether.  In this way, it is necessary to fully explore the implications of the new LBFS characterization.  Being specified in terms of LBFS end vertices, it appears uniquely suited to the incremental setting of this paper.  It remains to be seen if it can be applied to some benefit in the ``offline'' setting.  Linear time circle graph recognition via the LBFS characterization could still be a possibility with such an approach.

%DGC2 minor rewording below
But there may yet be additional applications of the incremental split decomposition algorithm coupled with the LBFS characterization.  One possibility for exploration is rank-width determination.  Its connection with circle graphs was noted in the introduction.  However, there are also connections with split decomposition.  For example, distance-hereditary graphs -- the family of graphs without prime subgraphs -- are precisely the graphs with rank-width 1.  
An algorithm to determine the split decomposition of distance-hereditary graphs appeared in \cite{GP07, GP08} using a restricted version of the 
algorithm presented in our companion paper.
It would be interesting to investigate what LBFS and split decomposition can together reveal about other graphs of bounded rank-width.
% EG 09/23/12 - added ref \cite{GP08} that deserves to be cited when mentioning DH graphs, but the paragraph above should be shortened, since the connection with rank-width is a rough vague idea, not much promising according to Christophe and I
%
% EG 09/23/12 - added keywords below following Christophe's intuition
%Other interesting connections might be: permutation graphs, parity graphs...
Similarly, could LBFS and split decomposition yield fast simple recognition algorithms for permutation graphs (strictly contained in circle graphs)
as well as parity graphs and Meyniel graphs? Both families strictly contain distance-hereditary graphs.

\bibliographystyle{plain}

%\bibliography{biblio-circle}

\end{document}